\RequirePackage[T1]{fontenc}
\documentclass[12pt]{article}
\pdfoutput=1 
\usepackage[height=8.85in,width=6.45in]{geometry}

\usepackage[utf8]{inputenc}
\usepackage{amsmath}
\usepackage{amssymb}
\usepackage{mathrsfs}
\usepackage{mathtools}
\numberwithin{equation}{section}
\usepackage[svgnames]{xcolor}
\usepackage[colorlinks,linktocpage=true,citecolor=DarkGreen,linkcolor=FireBrick]{hyperref}
\usepackage{cite}
\usepackage{graphicx}
\usepackage{tikz}
\usepackage{times}
\usepackage[scaled]{couriers}
\usepackage{subfig}
\usepackage{mathrsfs}
\usepackage{amsthm}

\usepackage{xcolor}
\usepackage{mdframed}

\renewenvironment{figure}[1][]{
  \begin{originalfigure}[#1]
    \begin{mdframed}[linecolor=black!0,backgroundcolor=black!1]
}{
    \end{mdframed}
  \end{originalfigure}
}

\usepackage{kazuya}

\newcommand{\Stueckelberg}{{Stueckelberg} } 
\newcommand{\A}[1]{A_{(#1)}}
\newcommand{\B}[1]{B_{(#1)}}

\newcommand{\act}[1]{\,\sp{#1}\!}
\newcommand{\actwedge}{{\wedge^\triangleright}}
\newcommand{\pwedge}{{\wedge^{\{,\}}}}
\newcommand{\HLawedge}{{\wedge^{\text{\tiny $HL$}}}}
\newcommand{\HLbwedge}{{\wedge^{\text{\tiny $HL'$}}}}
\newcommand{\liewedge}{{\wedge^{[,]}}}

\newcommand{\scA}{\mathscr{A}}

\newcommand{\fr}{\frac}

\begin{document}

\begin{titlepage}

\begin{flushright}
TU-1297
\end{flushright}

\vskip 3cm

\begin{center}

{\Large \bfseries 3-Crossed Module Structure in the Five-Dimensional Topological Axion Electrodynamics}

\vskip 1cm
Masaki Fukuda${}^1$,
Tommy Shu${}^2$
and 
Ryo Yokokura${}^3$
\vskip 1cm

\begin{tabular}{ll}
${}^1$Department of Physics, Tohoku University, Sendai 980-8578, Japan \\
${}^2$Department of Mathematics, Tohoku University, Sendai 980-8578, Japan \\
${}^3$Department of Physics \& Research and Education Center for Natural Sciences,\\ \hphantom{${}^3$}Keio University, Hiyoshi 4-1-1, Yokohama, Kanagawa 223-8521, Japan
\end{tabular}

\vskip 1cm

\end{center}

\noindent
In this paper, we investigate the higher-group symmetry structure of a five-dimensional topological theory, which is described by a 3-crossed module. The model is obtained by a five-dimensional extension of topological axion electrodynamics in four dimensions. To study the symmetry structure, we couple background gauge fields to the symmetry currents via Stueckelberg couplings. We show that background gauge invariance requires modified gauge transformation laws, indicating the existence of a higher-group structure. Furthermore, we identify the underlying mathematical structure as a 3-crossed module by regarding the modified Stueckelberg couplings as curvatures of a higher-group gauge theory. We demonstrate that the gauge transformation laws derived from this algebraic structure are consistent with the analysis based on the gauge invariance. 
While our previous work introduced the concept of a 3-crossed module motivated by higher-group symmetries, this work provides concrete verification that this framework correctly captures the symmetry structure of physical theories.

\end{titlepage}

\setcounter{tocdepth}{2}

\newpage

\tableofcontents

\section{Introduction}
Investigating global symmetries of a given theory is an important first step when studying a system described by that theory.
\footnote{In this paper, we will refer to global symmetries as ``symmetries'' and refer to gauge symmetries as ``gauge invariance.''} 
They provide valuable insights for understanding the system, and in some cases, they even allow us to solve it completely. 

Recently, the notion of the symmetries in quantum field theory has been refined in terms of topological operators, revealing the existence of various types of symmetries \cite{Gaiotto:2014kfa}. Their new interpretation can be summarized as follows. Suppose that an operator is inserted in the spacetime; It may be localized at a point or extended along some spatial or temporal directions. If the correlation functions remain invariant under continuous deformations of this operator, we say that the operator is \textit{topological} and the theory possesses a symmetry associated with it. In other words, the existence of a symmetry is reinterpreted as the existence of a topological operator. The new symmetries are referred to as generalized symmetries. Such symmetries are not only of mathematical interest but also play essential roles in a wide range of physical contexts, from condensed matter systems to quantum gravity. 

Higher-form symmetries are typical examples of the generalized symmetries whose symmetry operators act on extended objects along certain spatial directions. In particular, a symmetry whose operators act on $p$-dimensional objects is called a $p$-form symmetry. In terms of this convention, the ordinary symmetries are referred to as 0-form symmetries. Such symmetries, for example, allow us to understand topological order, which cannot be characterized by the order parameters of 0-form symmetries \cite{Thouless:1982zz,Laughlin:1983fy,Wen:1989iv,Wen:1990zza,Wen:1991rp}, in terms of symmetry breaking of higher-form symmetries\cite{Nussinov:2006iva,Nussinov:2009zz,Gaiotto:2014kfa}.\footnote{Non-invertible symmetries have also been actively studied in recent years, and provide novel insights and tools for understanding various physical phenomena. However, we will not discuss them further in this paper. For more details, see \cite{McGreevy:2022oyu,Cordova:2022ruw,Brennan:2023mmt,Shao:2023gho,Luo:2023ive,Gomes:2023ahz,Copetti:2023mcq} and references therein. In particular, non-invertible symmetries in axion electrodynamics have been discussed in \cite{Yokokura:2022alv,Choi:2022fgx,DelZotto:2024ngj}.} Much like ordinary symmetries, these generalized symmetries are becoming indispensable tools for understanding fundamental aspects of physics, and investigating them is also an essential first step when studying a given theory.

In this paper, we focus on \textit{higher group symmetries}\cite{Sharpe:2015mja,Cordova:2018cvg,Benini:2018reh}, in which topological operators of different dimensions are correlated with each other. Equivalently, one can say that symmetry operators themselves carry charges under other symmetry operators. A useful way to analyze such a situation is through background gauging. This is a procedure that promotes a symmetry transformation parameter to a function of spacetime and introduces a gauge field so that the theory remains invariant under the deformed transformation. Here, the gauge fields introduced in this procedure are treated as external fields. Since these fields can be regarded as networks of the topological defects associated with the symmetries \cite{Gaiotto:2014kfa}, we can use them to study the properties of the corresponding symmetries. In the presence of higher-group symmetries, gauge transformation laws and Bianchi identities can be modified due to the above interrelations among symmetries. Thus, higher-group structures are encoded in these deformations. 

Although deformations of gauge transformations and Bianchi identities signal the presence of higher-group symmetries, the precise mathematical structures underlying them are not always well understood. For instance, in higher dimensional extensions of axion electrodynamics \cite{Nakajima:2022feg,Nakajima:2024vgc}, it is expected that the interplay between electric and magnetic symmetries should be described by higher \textit{crossed modules}, one of the mathematical formulations of higher-groups. Yet the precise mathematical structure has not been fully understood, since a satisfactory definition of the gauge fields and their gauge transformation laws for higher crossed modules beyond the 2-crossed module (the three-group) is still a subject of active discussion.

In this paper, we focus on such situations and attempt to clarify the underlying structures. In the previous work by two of the present authors (MF, TS) \cite{Fukuda2025}, inspired by generalized global symmetries and higher-dimensional extensions of Dijkgraaf-Witten theory \cite{Girelli:2007tt,Radenkovic:2019qme} (see also subsequent developments), a categorical formulation of a 3-crossed module has been introduced.
\footnote{
Regarding the definition of 3-crossed modules (and their associated 4-groups), see, for example, the pioneering work \cite{zbMATH05664819}. Their approach is based on the Moore complex of a simplicial group. Specifically, they demonstrated that the category of 3-crossed modules is equivalent to that of Moore complexes of length three. This result is significant as it translates combinational-topological data into a computable algebraic framework. Meanwhile, our definition is also motivated by the aim to generalize Gray-categories to the level of 4-categories. A more detailed treatment of these categorical aspects will be provided in \cite{Fukuda:2026}.
} 
Naively, one may expect that the higher-group symmetry of higher-dimensional axion electrodynamics can be described in terms of this object. However, even in four dimensions, nontrivial 2-crossed module operations already are required \cite{Hidaka:2021kkf,Hidaka:2021mml}, and in dimensions higher than six, the relevant crossed module structure becomes increasingly intricate. For this reason, we begin by constructing a five-dimensional toy model, which is expected to exhibit a higher-group symmetry structure, and show that its higher-group symmetry can indeed be described in terms of the 3-crossed module. 

The structure of this paper is as follows. In Sec.~\ref{sec:action}, we construct a toy model as a higher-dimensional extension of topological axion electrodynamics (TAE), with particular emphasis on electric and magnetic symmetries. In Sec.~\ref{sec:BGgauge}, we turn on the background gauge fields, compute the 't~Hooft anomalies, and analyze the resulting symmetry structure in more detail. In Sec.~\ref{sec:str}, we reformulate the background gauge fields in terms of four-group gauge theory and extract the underlying 3-crossed module structure. Finally, in Sec.~\ref{sec:summary-discussion}, we summarize our results. 

Since the axioms of a 3-crossed module are somewhat intricate, we present them only in App.~\ref{CM}. In addition, App.~\ref{App:FG} contains a summary of four-group gauge theory, which we employ to analyze the mathematical structure.

\section{Five-Dimensional Topological Axion Electrodynamics\label{sec:action}}
In this section, we extend the TAE \cite{Hidaka:2021kkf,Hidaka:2021mml} to a five-dimensional theory. As mentioned in the Introduction, we are interested in the global symmetries of the theory since they provide important hints to analyze the system.

The $2n$-dimensional extensions of Axion electrodynamics have been studied in \cite{Nakajima:2022feg,Nakajima:2024vgc}. As anticipated from these works, electric and magnetic symmetries$-$constructed from the equations of motion and the Bianchi identities$-$in our theory can be elegantly described in terms of a 3-crossed module. To investigate this structure, we will introduce topological operators associated with these symmetries. 

\subsection{Action} 
In this section, we present an action with a symmetry structure described by a nontrivial 3-crossed module. To construct such a theory, we follow two strategies. Since topological operators of different codimensions can correlate with one another in a theory with higher-group symmetry, as discussed in the Introduction, it is natural to expect a higher-group structure to appear in a theory with rich higher-form symmetries. A higher dimensional theory is therefore a suitable setting for our purpose, as it allows for various kinds of topological operators. 

Furthermore, we will consider a topological theory. Typically, a topological theory, such as a BF theory, arises in the low energy limit of a gapped theory. Since all local excitations are gapped out and correlation lengths become effectively infinite, any operator in such a theory becomes topological. Consequently, topological theories exhibit a large number of symmetries. For example, in the low energy limit of a gapped Abelian Higgs model, where local particle propagation and the width of topological solitons can be neglected, the effective theory is described by a BF like theory known as TAE \cite{Hidaka:2021mml,Hidaka:2021kkf}. In such an IR regime, various higher-form symmetries associated with the conservation of the number of topological solitons are expected to emerge. Indeed, TAE exhibits a higher-group structure characterized by a nontrivial 2-crossed module. For these reasons, we aim to extend the theory to five dimensions. Naively, one expects that such a topological theory exhibits a higher-group structure characterized by a 3-crossed module. 

Higher dimensional generalizations of TAE can be constructed
as follows. Let $M_d$ be a $d$-dimensional oriented and closed manifold, which we regard as spacetime. We introduce $U(1)\ i$-form gauge fields $a_{(i)},\ (i=0,1,2,\cdots,d-1)$ together with their field strengths $\diff a_{(i)}$. We assume that they satisfy the Dirac quantization conditions:
\begin{equation}
  \label{eq:DiracQ}
  \int_{\cC_i}\frac{\diff a_{(i)}}{2\pi} \in \bZ,
\end{equation}
and transform under gauge transformations as:
\begin{equation}
 a_{(i)} \to a_{(i)} + \left\{\begin{array}{ll}\diff\theta_{(i-1)} & i=1,2,\cdots,d-1,\\ 2\pi & i=0,\end{array}\right.
\end{equation}
where $\cC_i,\ (i=0,1,2,\cdots,d-1)$ are cycles of $M_d$, and $\theta_{(i-1)}$ are gauge transformation parameters subject to $\int_{\cC_i}\diff\theta_{(i)}/2\pi\in\bZ$. The factor of $2\pi$ in the denominator is a matter of convention. In this convention, the Wilson loop operators take a particularly simple form: They are given by the exponential of the gauge fields integrated over a closed submanifold of spacetime, without any additional numerical coefficients. By taking wedge products of these field strengths, we obtain various exact forms such as $\diff a_{(i)}\wedge\diff a_{(j)}\wedge \ldots$. These are referred to as Chern-Weil currents~\cite{Heidenreich:2020pkc,Brauner:2020rtz}. We then couple the $U(1)$ higher-form gauge fields to the Chern-Weil currents so that the resulting differential forms have degree $d$. The resulting terms are topological, in the sense that they do not depend on the spacetime metric. The action is constructed as an integer linear combination of these topological terms. Further, we include generalizations of BF terms for the dynamical gauge fields like $a_{(3)}\wedge \diff a_{(1)}$ as topological mass terms to realize a gapped theory.

Following this procedure, we obtain a five-dimensional extension of TAE:
\begin{align}
 \nonumber
 S[M_5] &= \int_{M_5}\frac{p_4}{2\pi}a_{(4)}\diff a_{(0)} + \frac{p_3}{2\pi}a_{(3)}\diff a_{(1)} + \frac{p_2}{2\pi}\tilde a_{(2)}\diff a_{(2)} \\
  \label{eq:Action}
    &\hspace{1cm} + \frac{1}{(2\pi)^2}\vec p_2\cdot \vec a_{(2)}\diff a_{(0)}\diff a_{(1)} + \frac{p_1}{(2\pi)^2}a_{(1)}\diff a_{(1)}{}^2,
\end{align}
where $p_i, (i=1,2,3,4)$ are integers, $\vec p_2 = (p_{2,1},p_{2,2})$ are integers and $\vec a_{(2)} = (a_{(2)},\tilde a_{(2)})$ denotes a pair of 2-forms. The terms in the first line of \eqref{eq:Action} are BF terms which realize the gapped theory.
The gauge fields $a_{(0)}$ and $a_{(1)}$ can be regarded as the axion and the photon, and $a_{(4)}$ and $a_{(3)}$ are gauge fields which give masses for the axion and photon, respectively. The terms in the second line are topological interaction terms, which are higher dimensional analogues of the axion-photon coupling. In five dimensions, the photon can have the Chern-Simons coupling 
as introduced in the last term. The gauge fields $a_{(0)}$ and $a_{(1)}$ can be topologically interacted through 2-form gauge fields, which can be described by 
the first term in the second line. The topological mass term of the 2-form 
gauge fields can be introduced by $a_{(2)} \diff \tilde{a}_{(2)}$. Throughout, we omit the wedge product symbol $\wedge$ unless its inclusion is necessary to avoid confusion. Here, we do not include terms such as $a_{(2)} \diff a_{(2)}$, 
because they vanish up to a total derivative. As we mentioned before, the spacetime manifold $M_5$ is oriented and closed.

Before going to the next subsection, we comment on 
the topological mass term  $a_{(2)} \diff \tilde{a}_{(2)}$.
One may consider the action without the mass term,
\begin{equation}
  \label{eq:withoutTilde}
 S_{\rm 0}[M_5] = \int_{M_5}\frac{p_4}{2\pi}a_{(4)}\diff a_{(0)} + \frac{p_3}{2\pi}a_{(3)}\diff a_{(1)} + \frac{p_2}{(2\pi)^2}a_{(2)}\diff a_{(0)}\diff a_{(1)} + \frac{p_1}{(2\pi)^2}a_{1}\diff a_{(1)}{}^2
\end{equation}
However, in this case, the degrees of freedom of the 2-form field $a_{(2)}$ can be translated into those of a 1-form field by performing a dualization. To see this, let us consider the following simple action that includes a kinetic term:
\begin{equation}
  \label{eq:pice}
 S_{\rm kin.} = \int_{M_5}-\frac{1}{2g^2}\diff a_{(2)}\wedge*\diff a_{(2)} + \frac{iN}{2\pi}a_{(0)}\diff a_{(1)}\wedge\diff a_{(2)},
\end{equation}
where $g$ is a coupling constant, $N$ is an integer and the asterisk $*$ denotes the Hodge star operation defined by the metric on $M_5$. Irrelevant terms in the action \eqref{eq:withoutTilde} are omitted for simplicity. Taking the strong coupling limit $g^2\to \infty$, the system reduces to the original theory \eqref{eq:withoutTilde}. \\%
\indent%
In order to replace the field strength $F_{(3)}=\diff a_{(2)}$ by a dual 1-form field $\tilde a_{(1)}$, defined as the potensial of the dual field strength $\tilde F_{(2)} = *F_{(3)}$, let us introduce a delta functional term as follows:
\begin{equation}
  \label{eq:med}
 S_{\rm dual} = \int_{M_5}-\frac{1}{2g^2}F_{(3)}\wedge*F_{(3)} + \frac{iN}{2\pi}a_{(0)}\diff a_{(1)}\wedge\diff a_{(2)} + \frac{i\tilde F_{(2)}}{2\pi}\wedge\left(F_{(3)} - \diff a_{(2)}\right).
\end{equation}
In this expression, both of the 2-form field $a_{(2)}$ and its strength $F_{(3)}$ are treated as independent variables, while the new 2-form field $\tilde F_{(2)}$ plays the role of a Lagrange multiplier. By integrating out the original fields $F_{(3)}$ and $a_{(2)}$, we can change the variables of the theory. To perform this procedure, we first derive the equations of motion for these fields. For the field strength $F_{(3)}$, we obtain
\begin{equation}
  \label{eq:eomF}
 F_{(3)} = \frac{ig^2*\tilde F_{(2)}}{2\pi}.
\end{equation}
Physically, this corresponds to an exchange between the electric and magnetic fields in the system. Next, for the field $a_{(2)}$, we obtain a condition stating that the combination $\tilde F_{(2)} - Na_{(0)}\diff a_{(1)}$ is a closed form. Therefore, locally it can be written as
\begin{equation}
  \label{eq:eomb}
 \tilde F_{(2)} = \diff \tilde a_{(1)} + Na_{(0)}\diff a_{(1)},
\end{equation}
where we have introduced a dual 1-form field $\tilde a_{(1)}$. Substituting Eqs. \eqref{eq:eomF} and \eqref{eq:eomb} into the action \eqref{eq:med}, and using Stokes's theorem, we obtain
\begin{equation}
 S_{\rm kin.,dual} = \int_{M_5}-\frac{g^2}{8\pi^2}\tilde F_{(2)}\wedge*\tilde F_{(2)},
\end{equation}
where the dual field strength $\tilde F_{(2)}$ is given by \eqref{eq:eomb}. This action is locally equivalent to the original one, Eq.~\eqref{eq:pice}. 

Thus, in this case, the operator algebra tends to simplify because the degree of the differential form is reduced through this dualization procedure. Furthermore, taking the limit $g^2\to \infty$, which reduces the theory with kinetic term \eqref{eq:pice} to the original one \eqref{eq:withoutTilde}, the dual field strength must vanish, i.e. $\tilde F_{(2)} = 0$, further simplifying the original theory \eqref{eq:withoutTilde} and eventually leaving two decoupled BF-like theories. Such cases have been well studied in the literature \cite{Horowitz:1989ng,Blau:1989dh,Blau:1989bq,Birmingham:1991ty}.

To prevent such a simplification, we instead focus on situations where the 2-form fields cannot be dualized to 1-forms. The simplest way to realize this is to introduce an additional 2-form field and couple it through a mixing term. Denoting these fields by $a_{(2)}$ and $\tilde a_{(2)}$, one can add a BF-like mixing term of the form $\tilde a_{(2)}\wedge\diff a_{(2)}$. The presence of this term obstructs the dualization of the 2-form fields.

Such a situation, where two types of 2-form fields appear, arises in the context of string theory. Type IIB string theory contains two kinds of 2-form fields: the Kalb-Ramond field $B^{\text{\tiny NS}}$ and the Ramond-Ramond field $B^{\text{\tiny RR}}$. Let us now consider the IR theory, which is obtained as a low energy effective description of some UV completion. If we assume that the UV theory originates from type IIB string theory, the appearance of two distinct 2-form fields is quite natural. Furthermore, a mixing term of this kind appears in the symmetry topological field theory associated with the $\text{AdS}_5\times S^5$ setup \cite{Witten:1998wy}.\footnote{See also a reference \cite{Apruzzi:2021nmk} and related papers, for example \cite{Najjar:2024vmm,Najjar:2025htp,Khlaif:2025jnx,Bergman:2025isp}. If we take a good internal manifold, our theory might appear as a symTFT in a similar setup. } 

\subsection{Electric and Magnetic Symmetries}\label{sec:GS}
If we regard equations of motion and Bianchi identities as current conservation laws, then we can obtain various higher-form symmetries. They are called electric and magnetic symmetries, respectively. In addition to these symmetries, further conserved currents can be constructed by combining closed forms as long as their degrees do not exceed the spacetime manifold $M_d$. 
In particular, the symmetries obtained by combining magnetic symmetries are called Chern-Weil symmetries~\cite{Heidenreich:2020pkc,Brauner:2020rtz}. In this section, we will identify these symmetries. 

First, we now see electric symmetries. If we define the conserved currents as follows, then we can show that these currents are conserved by using the equations of motion:
\begin{align}
 J^E_{(0)} &= \frac{p_4}{2\pi}a_{(0)}, &
 J^E_{(1)} &= \frac{p_3}{2\pi}a_{(1)}, \\
 J^E_{(2)} &= \frac{p_2}{2\pi}\tilde a_{(2)} - \frac{p_{2,1}}{(2\pi)^2}a_{(0)}\diff a_{(1)}, &
 \tilde J^E_{(2)} &= \frac{p_2}{2\pi}a_{(2)} + \frac{p_{2,2}}{(2\pi)^2}a_{(0)}\diff a_{(1)},
\end{align}
\begin{align}
 J^E_{(3)} &= \frac{p_3}{2\pi}a_{(3)} + \frac{1}{(2\pi)^2}\vec p_2\cdot \vec a_{(2)}\diff a_{(0)} + \frac{3p_1}{(2\pi)^2}a_{(1)}\diff a_{(1)}, \\
 J^E_{(4)} &= \frac{p_4}{2\pi}a_{(4)} + \frac{1}{(2\pi)^2}\vec p_2\cdot \vec a_{(2)}\diff a_{(1)}.
\end{align}
We can construct electric $i$-form symmetry operators:
\begin{equation}
  \label{eq:EleSym}
 U_i(e^{i\alpha_i},\cC_{4-i}) = \exp\left(-i\alpha_i\int_{\cC_{4-i}}J^E_{(4-i)}\right),\qquad i=0,1,2,3,4,
\end{equation}
$\alpha_i$ are restricted to discrete values by the large gauge invariance. Let us explain this restriction more concretely. As an illustration, let us consider the case of the 0-form symmetry. When $\alpha_0$ takes arbitrary values in $[0,2\pi)$, the operator \eqref{eq:EleSym} is not generically gauge invariant. To define this operator in a gauge invariant way, we introduce a bulk auxiliary manifold $\cS_5$ whose boundary is the four cycle $\cC_4$, i.e. $\partial \cS_5 = \cC_4$. By Stokes' theorem, the integrand can be rewritten as $\diff J^E_{(4)}$, which in turn can be expressed in terms of the gauge-invariant forms $\diff a_{(i)}$:
\begin{equation}
 U_0(\cC_4) = \exp\left(-i\alpha_0\int_{\cS_5}\diff J^E_{(0)}\right)
\end{equation}
This construction makes the operator appear gauge invariant for arbitrary values of $\alpha_0$. However, an ambiguity remains, since the resulting topological operator depends on the choice of the auxiliary manifold. In other words, the definition is manifold dependent. To ensure both uniqueness and gauge invariance, $\alpha_0$ must be restricted to discrete values. To see this, let us impose the condition
\begin{equation}
  \label{eq:U0const}
 \exp\left(-i\alpha_0\int_{\cC_5}\diff J^E_{(4)}\right) = 1
\end{equation}
where $\cC_5 = \cS_5\cup \bar \cS_5'$ is a submanifold of the spacetime $M_5$, constructed by gluing $\cS_5$ and $\cS_5'$ along their common boundary $\cC_4$. Since
\begin{equation}
 N_4:=\int_{\cC_5}\frac{\diff a_{(4)}}{2\pi},\quad N_{2,1}:=\int_{\cC_5}\frac{\diff a_{(2)}}{2\pi}\frac{\diff a_{(1)}}{2\pi},\quad N_{2,2}:=\int_{\cC_5}\frac{\diff \tilde a_{(2)}}{2\pi}\frac{\diff a_{(1)}}{2\pi}
\end{equation}
take integer values because of Dirac quantization conditions \eqref{eq:DiracQ}. In terms of these integers, the integral can be rewritten as
\begin{equation}
  \int_{\cC_5}\diff J_E = P_0\left(p_4'N_4 + p_{2,1}'N_{2,1} + p_{2,2}'N_{2,2}\right),
\end{equation}
where $p_4',p_{2,1}'$ and $p_{2,2}'$ are coprime integers and $P_0:=\gcd(p_4,p_{2,1},p_{2,2})$ is a greatest common divider of $p_4,p_{2,1}$ and $p_{2,2}$. The strictest condition is therefore
\begin{equation}
  \label{eq:U0const2}
 e^{-i\alpha_0P_0} = 1.
\end{equation}
and condition \eqref{eq:U0const} is automatically satisfied whenever \eqref{eq:U0const2} holds. It follows that $\alpha_0$ must be quantized as
\begin{equation}
  \alpha_0 = 2\pi i\frac{n_0}{P_0},\qquad n_0 = 0,1,\cdots,P_0-1.
\end{equation}

By repeating the same procedure, we find that the parameters $\alpha_i$ are quantized to discrete values given by
\begin{equation}
  \alpha_i = 2\pi \frac{n_i}{P_i},\qquad n_i=0,1,\cdots,P_i - 1
\end{equation}
with
\begin{align}
 P_0&=\gcd(p_{2,1},p_{2,2},p_4), & P_1 &= \gcd(3p_1,p_{2,1},p_{2,2},p_3), & \left\{\begin{array}{c}P_2=\gcd(p_2,p_{2,1}),\\ \tilde P_2=\gcd(p_2,p_{2,2}),\end{array}\right. \\
 P_3&=p_3, & \quad P_4&=p_4.
\end{align}

Next, we consider the magnetic and Chern-Weil symmetries. There are five magnetic symmetries associated with the five fundamental fields:
\begin{align}
 V_i(e^{i\beta_i},\cC_{4-i}) = \exp\left(i\beta_i\int_{\cC_{4-i}}\frac{\diff a_{(3-i)}}{2\pi}\right),\qquad i=-1,0,1,2,3
\end{align}
where $e^{i\beta_i} \in U(1)$. In the present case, there is no restriction on the parameters $\beta_i$, because the conserved currents are gauge invariant on their own.
\footnote{If we have dynamical degrees of freedom with monopole charge, there are restrictions on parameters because they violate the Bianchi identity dynamically. In the present case, we have no restrictions because these dynamical degrees of freedom are absent.}

We also have Chern-Weil symmetries. The corresponding conserved currents are
\begin{align}
  \label{eq:CW1}
  &\frac{\diff a_{(0)}}{2\pi}\wedge\frac{\diff a_{(1)}}{2\pi}, &
  &\frac{\diff a_{(0)}}{2\pi}\wedge\frac{\diff a_{(2)}}{2\pi}, &
  &\frac{\diff a_{(0)}}{2\pi}\wedge\frac{\diff\tilde a_{(2)}}{2\pi}, \\
  \label{eq:CW2}
  &\left(\frac{\diff a_{(1)}}{2\pi}\right)^2, &
  &\frac{\diff a_{(1)}}{2\pi}\wedge\frac{\diff a_{(2)}}{2\pi}, &
  &\frac{\diff a_{(1)}}{2\pi}\wedge\frac{\diff\tilde a_{(2)}}{2\pi}, \\
  \label{eq:CW3}
  &\frac{\diff a_{(0)}}{2\pi}\wedge\frac{\diff a_{(3)}}{2\pi}.
\end{align}
Here as well, there are no restrictions on the parameters.

\section{Background Gauging\label{sec:BGgauge}}
It is straightforward to study correlations between topological defects directly. However, for simplicity, we employ a technique known as background gauging. This procedure promotes the global symmetry parameters to spacetime-dependent ones by introducing non-dynamical gauge fields that couple to the corresponding symmetry currents. Since this procedure corresponds to inserting a network of topological defects \cite{Gaiotto:2014kfa}, it provides an efficient field theoretic way to study their correlations. \\%
\indent%
In our setup, this gauging is implemented through \Stueckelberg couplings. However, the naive \Stueckelberg couplings must be appropriately modified to remove operator-valued ambiguities~\cite{Benini:2018reh}. This formulation allows us to extract the correlations between topological defects and the associated 't Hooft anomalies. In the next section, we will show that these gauge fields can be elegantly described in terms of a four-group gauge theory.

\subsection{Stueckelberg Coupling}
To make the gauge invariance manifest, as in the case of electric symmetries in Sec. \ref{sec:GS}, we work on an auxiliary six-dimensional manifold $X_6$ whose boundary is spacetime: $\partial X_6 = M_5$. Using Stokes theorem, the action becomes
\begin{align}
 \nonumber
 S[X_6] &= \int_{X_6}\frac{p_4}{2\pi}\diff a_{(4)}\diff a_{(0)} + \frac{p_3}{2\pi}\diff a_{(3)}\diff a_{(1)} + \frac{p_2}{2\pi}\diff\tilde a_{(2)}\diff a_{(2)} \\
    &\hspace{1cm} + \frac{1}{(2\pi)^2}\vec p_2\cdot \diff\vec a_{(2)}\diff a_{(0)}\diff a_{(1)} + \frac{p_1}{(2\pi)^2}\diff a_{(1)}{}^3.
\end{align}
This action has no ambiguity; in other words, it does not depend on the choice of the auxiliary manifold $X_6$. In this setup, the gauge invariance can be reinterpreted as independence on the choice of the auxiliary manifold. Therefore, this means that the original action \eqref{eq:Action} is invariant under the gauge transformation.

We now turn on the background gauge field associated with the electric symmetries. Roughly speaking, these symmetries act as shift symmetries. For example, a field $a_{(i)}$ transforms as $a_{(i)}\to a_{(i)} + \omega_{(i)}$, where $\omega_{(i)}$ is a closed $i$-form parameter. To gauge this symmetry, we promote the closed transformation parameter $\omega_{(i)}$ to be an arbitrary $i$-form, denoted by $\Lambda_{(i)}$. To maintain the invariance of the action under the transformation $a_{(i)}\to a_{(i)}+\Lambda_{(i)}$, we should introduce the background gauge field $A_{(i+1)}$ which transform as $A_{(i+1)} \to A_{(i+1)}+\diff \Lambda_{(i)}$, and implement the \Stueckelberg coupling \cite{Stueckelberg:1957zz} by replacing the field strength $\diff a_{(i)}$ with the invariant combination $\diff a_{(i)} - A_{(i+1)}$. As in ordinary 0-form symmetries and their background 1-form gauge fields,  the background gauge fields specify the places of 
the symmetry generators, and the gauge invariance 
for the background gauge transformations correspond to 
the conservation law.

The naive background gauging action is given by
\begin{align}
 \nonumber
 S_{\text{naive BG}}[X_6] &= \int_{X_6}\frac{p_4}{2\pi}(\diff a_{(4)}-\A5^{P_4})(\diff a_{(0)}-\A1^{P_0}) 
  + \frac{p_3}{2\pi}(\diff a_{(3)}-\A4^{P_3})(\diff a_{(1)}-\A2^{P_1}) \\
 \nonumber
  &\hspace{1.75em}
  + \frac{p_2}{2\pi}(\diff\tilde a_{(2)}-\A3^{\tilde P_2})(\diff a_{(2)}-\A3^{P_2}) \\
 \nonumber
  &\hspace{1.75em} + \frac{\vec p_2}{(2\pi)^2}\cdot(\diff \vec a_{(2)}-\vec{\A3})(\diff a_{(0)}-\A1^{P_0})(\diff a_{(1)}-\A2^{P_1}) \\
  \label{eq:naiveBG}
  &\hspace{1.75em} + \frac{p_1}{(2\pi)^2}(\diff a_{(1)}-\A2^{P_1})^3,
\end{align}
where $\A{i+1}^{P_i},\ (i=0,1,3,4)$ and $\vec{\A3} = (\A3^{P_2},\A3^{\tilde P_2})$ are $\bZ_{P_i}$ $i$-form symmetry background fields, which are flat gauge fields with $\bZ_{P_i}$-valued holonomies. These can be locally written in terms of Dirac quantized continuous gauge fields $\B{i}^{P_i}$ as follows:
\begin{equation}
  \label{eq:PA-dB}
 P_i\A{i+1}^{P_i} = \diff \B{i}^{P_i},\qquad \int_{\cC_{i+1}}\frac{\diff \B{i}^{P_i}}{2\pi} \in \bZ.
\end{equation}
As we will see below, the fields $\B{i}^{P_i}$ can be interpreted as background gauge fields for certain linear combinations of the magnetic and Chern-Weil symmetry currents. $\A{i+1}^{P_i}$ are constrained to $\bZ_{P_i}$ value holonomies by the above condition:
\begin{equation}
  \int_{\cC_{i+1}}\frac{\A{i+1}^{P_i}}{2\pi} \in \frac{\bZ}{P_i}.
\end{equation}
The action \eqref{eq:naiveBG} and \Stueckelberg coupling $\diff a_{(i)} - \A{i+1}^{P_i}$ are invariant under the following gauge transformation:
\begin{align}
  \label{eq:GTi}
 a_{(i)} &\to a_{(i)} + \Lambda_{(i)}, &
 \A{i+1}^{P_i} &\to \A{i+1}^{P_i} + \diff\Lambda_{(i)}, &
 \B{i}^{P_i} &\to \B{i}^{P_i} + P_i\Lambda_{(i)}.
\end{align}
Furthermore, since the continuous gauge fields $\B{i}^{P_i}$ are themselves gauge fields, the action is also invariant under their corresponding gauge transformations:
\begin{align}
  \label{eq:GTmag}
 \B{i}^{P_i} &\to \B{i}^{P_i} + \diff\lambda_{(i-1)} & i&\geq 1, \\
 \B{0}^{P_0} &\to \B{0}^{P_0} + 2\pi & i&=0.
\end{align}

However, the action \eqref{eq:naiveBG} suffers from ambiguities, in the sense that it depends on the choice of the auxiliary manifold $X_6$ as in the case of electric symmetries in Sec. \ref{sec:GS}. To see this, let us consider another auxiliary manifold $X_6'$ and construct a closed six-dimensional manifold $Z_6 = X_6\cup \bar X_6'$, where $\bar X_6'$ denotes $X_6'$ with reversed orientation, and the two manifolds are glued along the common boundary $M_5$. 

We can then evaluate $S_{\text{naive,BG}}[Z_6]$ as:
\begin{align}
 \nonumber
 S_{\text{naive,BG}}[Z_6] = &(\text{naive 't Hooft}) \\
 \nonumber
  &+ \int_{Z_6}\frac{p_{2,1}}{(2\pi)^2}\left(\diff a_{(2)}\A1^{P_0}\A2^{P_1}+\A3^{P_2}\diff a_{(0)}\A2^{P_1}+\A3^{P_2}\A1^{P_0}\diff a_{(1)}\right) \\
 \nonumber
  &\hspace{2em} + \frac{p_{2,2}}{(2\pi)^2}\left(\diff\tilde a_{(2)}\A1^{P_0}\A2^{P_1}+\A3^{\tilde P_2}\diff a_{(0)}\A2^{P_1}+\A3^{\tilde P_2}\A1^{P_0}\diff a_{(1)}\right) \\
 \nonumber
  &\hspace{2em} + \frac{3p_1}{(2\pi)^2}\diff a_{(1)}\A2^{P_1}\A2^{P_1} \mod 2\pi\bZ,
\end{align}
where $\text{(naive 't Hooft)}$ denotes an anomaly inflow action (with the opposite sign) associated with naive 't Hooft anomalies of the system, which will be determined below. These anomalies indicate that the partition function is not invariant under the background gauge transformation. This, however, poses no issue because 't Hooft anomalies simply reflect the obstructions to gauging the global symmetries. 

On the other hand, there exist ambiguities that depend on the dynamical fields $a_{(i)}$, known as operator-valued ambiguities~\cite{Benini:2018reh}. At first glance, these ambiguities resemble the inflow terms associated with 't Hooft anomalies. However, they are not 't Hooft anomalies since the corresponding action depends not only on the background fields but also on the dynamical ones. In other words, these ambiguities take values in $q$-numbers rather than $c$-numbers. Because of this feature, they may appear to lead to inconsistencies in the presence of topological defects. Nevertheless, as emphasized in \cite{Benini:2018reh}, such ambiguities are not true anomalies, since they can be cancelled by introducing appropriate couplings between the background and dynamical fields and by modifying the gauge transformation laws, in a manner analogous to the Green-Schwarz mechanism. The corresponding deformations of the \Stueckelberg coupling are
\begin{align}
 A^{P_4}_{(5)} &\to A^{\pi_4}_{(5)} = A^{P_4}_{(5)} + \frac{2\pi}{p_4}\left(\frac{p_{2,1}}{(2\pi)^2}A^{P_2}_{(3)}A^{P_1}_{(2)} + \frac{p_{2,2}}{(2\pi)^2}A^{\tilde P_2}_{(3)}A^{P_1}_{(2)}\right), \\
 A^{P_3}_{(4)} &\to A^{\pi_3}_{(4)} = A^{P_3}_{(4)} + \frac{2\pi}{p_3}\left(\frac{p_{2,1}}{(2\pi)^2}A^{P_2}_{(3)}A^{P_0}_{(1)} + \frac{p_{2,2}}{(2\pi)^2}A^{\tilde P_2}_{(3)}A^{P_0}_{(1)} + \frac{3p_1}{(2\pi)^2}A^{P_1}_{(2)}{}^2\right),
\end{align}
\vspace{-1em}
\begin{align}
 A^{P_2}_{(3)} &\to A^{\pi_2}_{(3)} = A^{P_2}_{(3)} + \frac{2\pi}{p_2}\cdot\frac{p_{2,2}}{(2\pi)^2}A^{P_0}_{(1)}A^{P_1}_{(2)}, &
 A^{\tilde P_2}_{(3)} &\to A^{\tilde\pi_2}_{(3)} = A^{\tilde P_2}_{(3)} - \frac{2\pi}{p_2}\cdot\frac{p_{2,1}}{(2\pi)^2}A^{P_0}_{(1)}A^{P_1}_{(2)}, \\
 A^{P_1}_{(2)} &\to A^{\pi_1}_{(2)} = A^{P_1}_{(2)}, &
 A^{P_0}_{(1)} &\to A^{\pi_0}_{(1)} = A^{P_0}_{(1)},
\end{align}
To see the connection with the underlying higher-group structure, it will be convenient to rewrite as follows. By using the relation \eqref{eq:PA-dB}, this can be expressed as
\begin{align}
  \label{eq:mod_BGF1}
 P_4A^{\pi_4}_{(5)} &= \diff \B4^{P_4} + \frac{p_{2,1}}{2\pi}A^{P_2}_{(3)}A^{P_1}_{(2)} + \frac{p_{2,2}}{2\pi}A^{\tilde P_2}_{(3)}A^{P_1}_{(2)}, \\
 P_3A^{\pi_3}_{(4)} &= \diff \B3^{P_3} + \frac{p_{2,1}}{2\pi}A^{P_2}_{(3)}A^{P_0}_{(1)} + \frac{p_{2,2}}{2\pi}A^{\tilde P_2}_{(3)}A^{P_0}_{(1)} + \frac{3p_1}{2\pi}A^{P_1}_{(2)}{}^2,
\end{align}
\vspace{-1em}
\begin{align}
 P_2A^{\pi_2}_{(3)} &= \diff \B2^{P_2} + \frac{p_{2,2}P_2}{2\pi p_2}A^{P_0}_{(1)}A^{P_1}_{(2)}, &
 \tilde P_2A^{\tilde\pi_2}_{(3)} &= \diff \B2^{\tilde P_2} - \frac{p_{2,1}\tilde P_2}{2\pi p_2}A^{P_0}_{(1)}A^{P_1}_{(2)}, \\
  \label{eq:mod_BGF2}
 P_1A^{\pi_1}_{(2)} &= \diff \B1^{P_1}, &
 P_0A^{\pi_0}_{(1)} &= \diff \B0^{P_0}.
\end{align}
This expression will be particularly useful as it can be naturally interpreted as a ``fake curvature condition'' for the modified background fields.

Note that all operator valued ambiguities can be eliminated by introducing the additional 2-form field $\tilde a_{(2)}$. The presence of this field enables us to add a Green-Schwarz-like term to the action, which absorbs the ambiguities. In contrast, without such additional degrees of freedom, as in the case of the action \eqref{eq:withoutTilde}, the ambiguities cannot be cancelled through the same mechanism.

As we mentioned in the following section, modification of gauge transformation laws is required by these modifications of the background gauge field couplings. This implies nontrivial correlations among topological operators of different dimensions.

By integrating over the closed submanifolds $\cC_i$, one finds that the holonomies are affected by the deformation, and that $\A{i+1}^{\pi_i}$ should be interpreted as $\bZ_{\pi_i}$ $i$-form symmetry background gauge fields, where $\pi_i$ are given by:
\begin{align}
  \pi_4 &= \frac{p_4P_1}{\gcd\left(P_1,p_{2,1}/P_2,p_{2,2}/\tilde P_2\right)}, &
  \pi_3 &= \frac{p_3P_0P_1}{\gcd\left(P_0P_1,p_{2,1}P_1/P_2,p_{2,2}P_1/\tilde P_2,3p_1P_0/P_1\right)}, \\
  \pi_2 &= \frac{p_2P_0}{\gcd\left(p_2P_0/P_2,p_{2,2}/P_1\right)},& \tilde\pi_2 &= \frac{p_2P_0}{\gcd\left(p_2P_0/\tilde P_2,p_{2,1}/P_1\right)}, \\
  \pi_1&=P_1,& \pi_0 &=P_0.
\end{align}
The resulting action is then:
\begin{align}
 \nonumber
 S_{\text{E,BG}}[X_6] &= \int_{X_6}\frac{p_4}{2\pi}(\diff a_{(4)}-\A5^{\pi_4})(\diff a_{(0)}-\A1^{\pi_0})
  + \frac{p_3}{2\pi}(\diff a_{(3)}-\A4^{\pi_3})(\diff a_{(1)}-\A2^{\pi_1}) \\
 \nonumber
  &\hspace{1.75em}
  + \frac{p_2}{2\pi}(\diff\tilde a_{(2)}-\A3^{\tilde \pi_2})(\diff a_{(2)}-\A3^{\pi_2}) \\
 \nonumber
  &\hspace{1.75em} + \frac{\vec p_2}{(2\pi)^2}\cdot(\diff \vec a_{(2)}-\vec{\A3^\pi})(\diff a_{(0)}-\A1^{\pi_0})(\diff a_{(1)}-\A2^{\pi_1}) \\
  \label{eq:ElectricBG}
  &\hspace{1.75em} + \frac{p_1}{(2\pi)^2}(\diff a_{(1)}-\A2^{\pi_1})^3,
\end{align}
where $\vec{\A3^\pi} = (\A3^{\pi_2},\A3^{\tilde\pi_2})$. This action contains no operator-valued ambiguities, while the 't Hooft anomalies remain and are modified. The resulting bulk action is
\begin{align}
 \nonumber
 (\text{'t Hooft}) &= \int_{X_6} 
 \frac{p_4}{2\pi}
 \A5^{P_4}\A1^{P_0} + \frac{p_3}{2\pi}\A4^{P_3}\A2^{P_1} + \frac{p_2}{2\pi}\A3^{\tilde P_2}\A3^{P_2} \\
  &\qquad + \frac{2p_{2,1}}{(2\pi)^2}\A3^{P_2}\A2^{P_1}\A1^{P_0} + \frac{2p_{2,2}}{(2\pi)^2}\A3^{\tilde P_2}\A2^{P_1}\A1^{P_0} + \frac{2p_1}{(2\pi)^2}\A2^{P_1}{}^3
\end{align}
These anomalies correspond to ambiguities unless the following fractions are integers:
\begin{align}
  &\frac{1}{P_0}, & 
  &\frac{1}{P_1}, &
  &\frac{p_2}{\tilde P_2P_2}, &
  &\frac{2p_{2,1}}{P_0P_1P_2}, &
  &\frac{2p_{2,2}}{P_0P_1\tilde P_2}, &
  &\frac{2p_1}{P_1^3}.
\end{align}

\subsection{Remarks on Gauge Transformations}\label{sec:SRGT}
The gauge transformations are also deformed due to the modification of the \Stueckelberg couplings. In particular, the action is no longer invariant under the simultaneous gauge transformations \eqref{eq:GTi} for all degree $i$.\footnote{The action and \Stueckelberg coupling remain invariant under the gauge transformations \eqref{eq:GTmag} for the background gauge fields $\B{i}$.}

However, the action and the \Stueckelberg couplings $\diff a_{(i)} - \A{i+1}^{\pi_i}$ remain invariant under the gauge transformation \eqref{eq:GTi} for a fixed degree $i$, since the deformed \Stueckelberg coupling contains the original undeformed form as part of its structure. That is,
\begin{equation}
 \diff a_{(i)} - \A{i+1}^{\pi_i} = \diff a_{(i)} - \A{i+1}^{P_i} + (\cdots),
\end{equation}
where ellipsis represents additional terms depending on the $\A{j+1}^{P_j}$ for $j<i$. Then, the gauge transformations of higher-degree background gauge fields (such as $\A{i+2}$) are mixed with those of lower-degree fields. This indicates the existence of higher-group symmetry. 

For example, let us consider the case $i=2$. The \Stueckelberg couplings $\diff a_{(2)} - \A3^{\pi_3}$ and $\diff \tilde a_{(2)} - \A3^{\tilde \pi_3}$ are invariant under the gauge transformation \eqref{eq:GTi} with $i=2$. Under this transformation, the deformed background gauge field $\A4^{\pi_3}$ transforms as:
\begin{equation}
 \A4^{\pi_3} \to \A4^{\pi_3} + \Delta\A4^{P_3} + \frac{p_{2,1}}{2\pi p_3}\diff\Lambda_{(2)}\A1^{P_0} + \frac{p_{2,2}}{2\pi p_3}\diff\tilde\Lambda_{(2)}\A1^{P_0},
\end{equation}
where $\Delta\A4^{P_3}$ denotes the deformed gauge transformation of $\A4^{P_3}$. In order to maintain the invariance of the \Stueckelberg coupling $\diff a_{(3)} - \A4^{\pi_3}$, the following gauge transformation law is required:
\begin{equation}
  \label{eq:3.19}
 \A4^{P_3} \to \A4^{P_3} - \frac{p_{2,1}}{2\pi p_3}\diff\Lambda_{(2)}\A1^{P_0} - \frac{p_{2,2}}{2\pi p_3}\diff\tilde\Lambda_{(2)}\A1^{P_0}.
\end{equation}
Similaly, the transformation of $\A5^{P_4}$ can be derived from that of $\A5^{\pi_4}$ by requiring the gauge invariance of $\diff a_{(4)} - \A5^{\pi_4}$. 

Performing this procedure for all degrees $i=0,1,2,3,4$, one can obtain the full set of deformed gauge transformations. But now we do not determine the gauge transformation laws in this way. Instead, we will determine them systematically using the general framework of four-group gauge theory. As we will see later, the gauge transformation laws derived from the four-group gauge theory precisely reproduce those obtained from the above calculation. Incidentally, the above transformation corresponds to a part of the gauge transformation of the third kind. 

Physically, these modified gauge transformation laws indicate correlations among topological operators of different dimensions. This can be understood as follows. As mentioned at the beginning of this section, turning on the background gauge fields corresponds to introducing a network of topological defects, so their gauge transformations correspond to recombinations of these defects. Because the gauge transformation laws are mixed, each recombination is affected by the presence of other topological operators with different dimensions.

\subsection{Background Gauging of the Magnetic Symmetries}
In this section, we comment on the background gauging of the magnetic and Chern-Weil symmetries. We couple the background gauge fields for the magnetic and Chern-Weil symmetries. In the absence of the background gauge fields of the electric symmetries, 
the coupling term can be written in $X_6$ as
\begin{align}
  \nonumber
  & S_{\rm mag., CW}[X_6] \\
  \nonumber
  &=  \int_{X_6}- \fr{\diff a_{(0)}}{2\pi} \diff C_{(4)} + \fr{\diff a_{(1)}}{2\pi}  
\diff C_{(3)}
- \fr{\diff a_{(2)}}{2\pi}  \diff {C}_{(2)} 
- \fr{\diff\tilde{a}_{(2)}}{2\pi}  \diff\tilde{C}_{(2)} 
+ \frac{\diff a_{(3)}}{2\pi}  \diff C_{(1)}
- \frac{\diff a_{(4)}}{2\pi}  \diff C_{(0)}
\\
\nonumber
& \qquad 
- \frac{\diff a_{(0)}}{2\pi} \frac{\diff a_{(1)}}{2\pi}  \diff C^{ (0,1)}_{(2)}
+ \frac{\diff a_{(0)}}{2\pi} \frac{\diff a_{(2)}}{2\pi}  \diff C^{ (0,2)}_{(1)}
- \frac{\diff a_{(0)}}{2\pi}\frac{\diff\tilde a_{(2)}}{2\pi}  \diff C^{(0,\tilde{2})}_{(0)}
\\
& \qquad 
+ \left(\frac{\diff a_{(1)}}{2\pi}\right)^2  \diff C^{(1,1)}_{(1)}
- \frac{\diff a_{(1)}}{2\pi}\frac{\diff a_{(2)}}{2\pi}  \diff C^{(1,2)}_{(0)}
- \frac{\diff a_{(1)}}{2\pi}\frac{\diff\tilde a_{(2)}}{2\pi}  \diff C^{(1,\tilde{2})}_{(0)}
- \frac{\diff a_{(0)}}{2\pi} \frac{\diff a_{(3)}}{2\pi}  \diff C^{(0,3)}_{(0)}  .
\end{align}
Here, the background $i$-form gauge fields $C_{(i)}$ and $C^{(p,q)}_{(i)}$ with integers $(p,q)$ are $U(1)$ gauge fields with the gauge transformation laws $C_{(i)} \to C_{(i)} + \diff \Upsilon_{(i-1)} $ and $C^{(p,q)}_{(i)} \to C^{(p,q)}_{(i)} + \diff \Upsilon^{(p,q)}_{(i-1)}$ by the gauge parameters $\Upsilon_{(i-1)}$ and $\Upsilon^{(p,q)}_{(i-1)}$, which satisfy the quantization conditions $\int_{{\cal C}_{i+1}} \diff C_{(i)} $, $\int_{{\cal C}_{i+1}} \diff C^{(p,q)}_{(i)} \in 2\pi \mathbb{Z}$, respectively. Since we work on the bulk manifold $X_6$, these currents couple to the exterior derivatives of background gauge fields, e.g., $\diff C_{(i)}$.

The introduced background gauge fields $C_{(3)}$ and $C_{(4)}$ can be absorbed into $B_{(3)}$ and $B_{(4)}$, which were previously introduced for the electric symmetries. Turning on the background gauge fields for electric symmetries, the total background coupling then becomes
\begin{align}
  \nonumber
  S_{\text{E,BG}}[X_6] + S_{\rm mag., CW}[X_6] &= S[X_6] + (\text{'t Hooft}) + (\text{ohter background couplings}) \\
  \label{eq:ElectricBG2}
  &\quad + \int_{X_6}
  \frac{\diff a_{(0)}}{2\pi}\diff(\B4^{P_4}-C_{(4)})
  - \frac{\diff a_{(1)}}{2\pi}\diff(\B3^{P_3}-C_{(3)}).
\end{align}
By redefining $\B4^{P_4}-C_{(4)} \to \B4^{P_4}$ and $\B3^{P_3}-C_{(3)} \to \B3^{P_3}$, we can eliminate the background fields $C_{(4)}$ and $C_{(3)}$. This implies that introducing independent background gauge fields for the magnetic symmetry currents $\diff a_{(0)}$ and $\diff a_{(1)}$ is redundant. Note that these redefinitions preserve the Dirac quantization conditions for the background gauge fields. Other fields cannot be absorbed in the same manner because such redefinitions would break these quantization conditions. In particular, $C_{(0)},C_{(1)}$, $C_{(2)}$ and $C_{(i)}^{(p,q)}$ cannot be absorbed into the electric background gauge fields $B$ in general, however, since these fields are not related to the 3-crossed module structure we analyze in the following sections, we do not pursue their detailed treatment here. 

In the following sections, for simplicity, we only turn on and focus on the background gauge fields for the electric symmetries. This is because the background fields for the magnetic and CW symmetries would lead to further modifications of the Bianchi identities and gauge transformations. We leave the detailed analysis of higher gauge theories in the presence of these background fields for future work.

\section{3-Crossed Module Structure\label{sec:str}}

In this section, we aim to interpret the symmetry structure of our theory in terms of a 3-crossed module, which provides one of the formulations of the four-group. The system \eqref{eq:Action} exhibits six higher-form symmetries ranging from $(-1)$- to 4-form, with nontrivial correlations among them. In particular, we argue that the correlations among the 0-,1-,2-, and 4-form symmetries can be understood as the operations of a 3-crossed module.

To extract this 3-crossed module structure, we interpret deformations of background gauge fields and flat curvature conditions as the vainshing of fake curvature conditions in the framework of higher gauge theories \cite{Hitchin:1999fh,Breen:2001ie,Baez:2004in,Baez:2010ya,Martins:2009evc,Schreiber:2008kcv,Saemann:2013pca,Wang:2013dwa,Schreiber:2008S,Martins:2007uki,MARTINS20113309,Martins:2009evc}.\footnote{In case of the four-group (3-crossed modules), see App.~\ref{App:4GT}.} As discussed in these references, these conditions are precisely those required for defining higher holonomies. Just as in the ordinary gauge theory, where the curvature is defined as $F=\diff A + \frac{1}{2}A\liewedge A$, the fake curvatures in higher gauge theory are defined through algebraic operations of the 3-crossed module such as Lie brackets, Peiffer liftings, and left actions. Therefore, by examining their explicit forms, one can read off the nontrivial operations characterizing the 3-crossed module structure. Furthermore, we demonstrate that the gauge transformation laws, such as Equation \eqref{eq:3.19}, are elegantly reproduced within this framework, indicating that the invertible symmetry structure is indeed captured correctly by the 3-crossed module. 

\subsection{Brief Introduction to Four-Group Gauge Theory}\label{sec:BriefIntro}
In this section, we briefly introduce the mathematical framework necessary to examine the structure of background gauge fields. The background gauge fields discussed in Section 3 can be described by a four-group gauge theory based on a 3-crossed module. A 3-crossed module roughly consists of the following data:\footnote{For more details, see App. \ref{App:3CM} and \cite{Fukuda2025}.}
\begin{itemize}
  \item Four groups $G,H,L$ and $M$, which we also denote by $G_0 = G,G_1=H,G_2=L$ and $G_3=M$, respectively.
  \item Homomorphisms between these groups, $\partial_i:~G_{i+1}\to G_i$, called boundary maps. They are required to satisfy the nilpotency condition $\partial_1\circ\partial_2(l) = e_G,\ \forall l\in L$ and $\partial_2\circ\partial_3(m) = e_H,\ \forall m\in M$. The subscripts on $\partial$ are often omitted.
  \item Actions of lower-degree group $G_i$ on higher-degree group $G_j$($j>i$), denoted by $x\triangleright y$ or $\act{x}y$ for $x\in G_i,y\in G_j$. For $i=j$, this is interpreted as the adjoint action: $\act{x}y = xyx^{-1}$.
  \item Six types of maps called liftings, which map some lower groups to a higher group $\{\text{-},\text{-},\cdots\}$\\$:~G_{i_1}\times G_{i_2}\times\ldots \to G_j$. For example, there is a map $\{-,-\}:~L\times L\to H$ called the Peiffer lifting.
  \item Coherence conditions coming from the consistency between group actions and boundary maps. For example, the relation $\act{\partial h_z}(h_yh_x) = \act{\partial h_z}h_y\act{\partial h_z}h_x$ implies that the Peiffer lifting $\{h_z,h_yh_x\}$ is equal to $\{h_z,h_y\}\act{\act{\partial h_z}h_y}\{h_z,h_x\}$ up to \textit{twist} of $\partial_3(M)$. Here the term ``twist'' refers a factor that vanishes under the boundary map $\partial$. This is automatically satisfied by elements in $\partial_3(M)$ due to the nilpotency condition $\partial\circ\partial(m) = e_H$. The relations that holds for 2-crossed modules are deformed by this factor. 
\end{itemize}

To study four-group gauge theory, it is convenient to introduce a Lie algebra version of the 3-crossed module, which is essentially the linearization of the group-level data as in ordinary gauge theories:
\begin{itemize}
  \item Four Lie algebras $\mathfrak{g},\mathfrak{h},\mathfrak{l}$ and $\mathfrak{m}$ corresponding to $G,H,L$ and $M$.
  \item Linear maps between these Lie algebras, $\delta_i:~\mathfrak{g}_{i+1}\to \mathfrak{g}_i$, satisfying the nilpotency condition $\delta_{i-1}\circ\delta_i = 0$. 
  \item Actions of $g\in G$ on the Lie algebras $X\in \mathfrak{g}_i$(denoted $g\triangleright X=\act{g}X$), and actions of lower algebras $X\in \mathfrak{g}_i$ on higher algebras $Y\in \mathfrak{g}_j$(denoted $X\triangleright Y = \act{X}Y$). These are the generalizations of adjoint actions.
  \item Six types of multilinear maps (liftings) from some lower algebras to a higher algebra \\ $\{\text{-},\text{-},\cdots\}:~\mathfrak{g}_{i_1}\otimes\mathfrak{g}_{i_2}\otimes\ldots \to \mathfrak{g}_j$. 
  \item Coherence conditions obtained by linearizing the group-level coherence conditions. 
\end{itemize}

As discussed in \cite{Martins:2009evc}, three-group holonomies are defined in terms of 2-crossed modules, where the authors provided expressions for the curvatures and imposed a consistency condition known as \textit{fake flatness} or \textit{vanishing fake curvature condition}. In our case, we do not know the precise form of fake curvatures from that perspective, since we have not established a full definition of four-group holonomies based on 3-crossed modules. Nevertheless, we can deduce the form of higher curvatures by analogy with the three-group case. We define the higher gauge transformation laws such that the flatness condition is preserved, following the approach discussed in \cite{Wang:2013dwa}.

Using the 3-crossed module operations, we define the fake curvatures as
\begin{align}
  \label{eq:Omega1}
  \Omega_1 &= \diff \mathscr{A}_{(1)} + \frac{1}{2}\mathscr{A}_{(1)}\liewedge \mathscr{A}_{(1)} - \delta\mathscr{A}_{(2)}, \\
  \label{eq:Omega2}
  \Omega_2 &= \diff \mathscr{A}_{(2)} + \mathscr{A}_{(1)}\actwedge\mathscr{A}_{(2)} - \delta\mathscr{A}_{(3)}, \\
  \Omega_3 &= \diff \mathscr{A}_{(3)} + \mathscr{A}_{(1)}\actwedge\mathscr{A}_{(3)} + \mathscr{A}_{(2)}\pwedge\mathscr{A}_{(2)} - \delta\mathscr{A}_{(4)}, \\
  \label{eq:Omega4}
  \Omega_4 &= \diff \mathscr{A}_{(4)} + \mathscr{A}_{(1)}\actwedge\mathscr{A}_{(4)} + \mathscr{A}_{(2)}\HLawedge\mathscr{A}_{(3)} + \mathscr{A}_{(2)}\HLbwedge\mathscr{A}_{(3)},
\end{align}
where $\mathscr{A}_{(i)},\ i=1,2,3,4$ are $i$-form gauge fields valued in $\mathfrak{g}_i$. Here, for differential $i$- and $j$- forms $\omega_{(i)}$ and $\eta_{(j)}$ valued in the Lie algebras $\mathfrak{g}$ and $\mathfrak{g}'$, respectively, the wedge product deformed by an operation $f: \mathfrak{g} \otimes \mathfrak{g}' \to \mathfrak{g}''$ is defined as follows:
\begin{equation}
  \omega_{(i)}\wedge^f\eta_{(j)} = \omega^\alpha_{(i)}\wedge \eta^A_{(j)}f(T_\alpha,T_A),\qquad T_\alpha\in\mathfrak{g},\quad T_A\in\mathfrak{g}'.
\end{equation}
In Eqs. \eqref{eq:Omega2}--\eqref{eq:Omega4}, the first two terms can be interpreted as a generalization of the covariant derivative, while the remaining terms arise from liftings.

Following the same computations as for $\Omega_2$ below, we obtain four types of transformation laws, including ordinary gauge transformation associated with the adjoint action. These are summarized in the App. \ref{App:4GT}. For example, a 4-gauge transformation of the third kind is
\begin{align}
  \label{eq:ThirdKind1}
  \mathscr{A}_{(1)} \to \mathscr{A}_{(1)}' &= \mathscr{A}_{(1)}, \\
  \mathscr{A}_{(2)} \to \mathscr{A}_{(2)}' &= \mathscr{A}_{(2)} + \delta\lambda, \\
  \mathscr{A}_{(3)} \to \mathscr{A}_{(3)}' &= \mathscr{A}_{(3)} + \diff\lambda + \mathscr{A}_{(1)}\actwedge\lambda, \\
  \label{eq:ThirdKind2}
  \mathscr{A}_{(4)} \to \mathscr{A}_{(4)}' &= \mathscr{A}_{(4)} - \mathscr{A}_{(2)}\HLawedge\lambda - \mathscr{A}_{(2)}'\HLbwedge\lambda - \lambda\pwedge\lambda,
\end{align}
where $\lambda$ is an $\mathfrak{l}$-valued 2-form gauge transformation parameter that shifts the 2-form gauge field $\mathscr{A}_{(2)}$ by the twist $\delta\lambda$.

Indeed, one can verify that the vanishing fake curvature conditions are preserved under these gauge transformations. To see this, let us examine how the fake curvatures transform under these transformations. Taking $\Omega_2$ as an example and substituting the transformation \eqref{eq:ThirdKind1}--\eqref{eq:ThirdKind2} into \eqref{eq:Omega2}, we obtain
\begin{equation}
  \label{eq:dummy1}
  \Omega_2 \to \Omega_2 + \delta(\mathscr{A}_{(3)} + \diff\lambda + \mathscr{A}_{(1)}\actwedge\lambda - \mathscr{A}_{(3)}')
\end{equation}
Equation \eqref{eq:dummy1} determines the transformation law for $\mathscr{A}_{(3)}$ such that the terms within the boundary map $\delta$ vanish. Following a similar procedure, we can determine the transformation laws for the 4-form gauge field $\mathscr{A}_{(4)}$ and for fake curvatures \eqref{eq:Omega1}--\eqref{eq:Omega4} as
\begin{align}
  \Omega_1 &\to \Omega_1, \\
  \Omega_2 &\to \Omega_2, \\
  \Omega_3 &\to \Omega_3 + \Omega_1\actwedge\lambda, \\
  \Omega_4 &\to \Omega_4 - \Omega_2\HLawedge\lambda - \Omega_2'\HLbwedge\lambda,
\end{align}
Consequently, if the fake curvatures vanish initially, they remain zero under these transformations. 

\subsection{Background Fields as Four-Group Gauge Theory}\label{sec:BFFGT}
To extract the higher-group structure underlying the action \eqref{eq:Action}, we combine the background gauge fields introduced in the previous section as follows:
\begin{align}
 \scA_{(1)} &= (\A1^{\pi_0},\B1^{P_1}) & G &= \bZ_{\pi_0}\times U(1) \\
 \scA_{(2)} &= (\A2^{\pi_1},\B2^{P_2},\B2^{\tilde P_2}) & H &= \bZ_{\pi_1}\times U(1)\times U(1) \\
 \scA_{(3)} &= (\A3^{\pi_2},\A3^{\tilde \pi_2},\B3^{\tilde P_3}) & L &= \bZ_{\pi_2}\times \bZ_{\tilde\pi_2}\times U(1) \\
 \scA_{(4)} &= (\A4^{\pi_3},\B4^{P_4}) & M &= \bZ_{\pi_3}\times U(1) \\
 \scA_{(5)} &= \A5^{\pi_4} & N &= \bZ_{\pi_4}
\end{align}
where $\scA_{(i+1)}$ are $(i+1)$-form valued in the Lie algebras of the groups on the right-hand side, respectively. We aim to rewrite relations between background gauge fields in the above section in terms of the four-group gauge theory given in App.~\ref{App:4GT}. To achieve this, we define boundary maps, left actions, and liftings\footnote{By a \textit{lifting} we mean a map from lower groups to a higher one. In the case of a 3-crossed module, there are six such maps, called Peiffer liftings and Homanians,} that characterize the 3-crossed module as follows:
\begin{itemize}
  \item Boundary maps:
  \begin{align}
    \label{eq:deltaA2}
    \delta_1\scA_{(2)} &= (0,P_1\A2^{\pi_1}) \\
    \delta_2\scA_{(3)} &= (0,P_2A^{\pi_2}_{(3)},\tilde P_2A^{\tilde\pi_2}_{(3)}) \\
    \delta_3\scA_{(4)} &= (0,0,P_3A^{\pi_3}_{(4)}) \\
    \label{eq:deltaA5}
    \delta_4\scA_{(5)} &= (0,P_4\A5^{\pi_4})
  \end{align}
  \item Left actions:
  \begin{align}
    \label{eq:A1actA2}
 \scA_{(1)}\actwedge\scA_{(2)} &= \left(0,\frac{p_{2,2}P_2}{2\pi p_2}A^{\pi_0}_{(1)}A^{\pi_1}_{(2)},-\frac{p_{2,1}\tilde P_2}{2\pi p_2}A^{\pi_0}_{(1)}A^{\pi_1}_{(2)}\right) \\
    \label{eq:A1actA3}
 \scA_{(1)}\actwedge\scA_{(3)} &= \left(0,0,\frac{p_{2,1}}{2\pi}A^{\pi_2}_{(3)}A^{\pi_0}_{(1)} + \frac{p_{2,2}}{2\pi}A^{\tilde \pi_2}_{(3)}A^{\pi_0}_{(1)}\right) \\
 \scA_{(1)}\actwedge\scA_{(4)} &= (0,0)
  \end{align}
  \item Liftings:
  \begin{align}
 \scA_{(2)}\pwedge\scA_{(2)} &= \left(0,0,\frac{3p_1}{2\pi}A^{\pi_1}_{(2)}{}^2\right) \\
 \scA_{(2)}\HLbwedge\scA_{(3)} &= \left(0,\frac{p_{2,1}}{2\pi }A^{\pi_2}_{(3)}A^{\pi_1}_{(2)} + \frac{p_{2,2}}{2\pi}A^{\tilde \pi_2}_{(3)}A^{\pi_1}_{(2)}\right) \\
    \label{eq:A2HLaA3}
 \scA_{(2)}\HLawedge\scA_{(3)} &= \left(0,0\right)
  \end{align}
\end{itemize}
Then, flat curvature conditions $\diff \A{i+1}^{\pi_i}=0$ and modified background gauge field $\A{i+1}^{\pi_i}$ \eqref{eq:mod_BGF1}--\eqref{eq:mod_BGF2} can be rewritten in terms of the vanishing fake curvature condition in the four-group gauge theory. 
\begin{align}
 \diff\scA_{(1)} &= \delta_1\scA_{(2)} \\
 \diff\scA_{(2)} + \scA_{(1)}\actwedge\scA_{(2)} &= \delta_2\scA_{(3)} \\
 \diff\scA_{(3)} + \scA_{(1)}\actwedge\scA_{(3)} + \scA_{(2)}\pwedge\scA_{(2)} &= \delta_3\scA_{(4)} \\
 \diff\scA_{(4)} + \scA_{(1)}\actwedge\scA_{(4)} + \scA_{(2)}\HLawedge\scA_{(3)} + \scA_{(2)}\HLbwedge\scA_{(3)} &= \delta_4\scA_{(5)}
\end{align}

From the above definitions, one finds that the symmetry analyzed in Sec.~\ref{sec:GS} can be classified in terms of the following 3-crossed module. 
As introduced in App.~\ref{App:3CM}, a 3-crossed module is a complex of four-groups, equipped with boundary maps, left actions of lower groups on higher ones, and liftings that satisfy axioms in App.~\ref{App:3CM}. In the present case, the complex involves five groups $N,\ M,\ L,\ H$, and $G$, which are connected via boundary maps.\footnote{If we also include $(-1)$-form symmetry, then six groups are connected.} However, only the liftings characteristic of a 3-crossed module appear explicitly in our theory. The liftings specific to a 4-crossed module do not arise in this setting.
Concretely, for the elements of the groups:
\begin{align}
 g &= (e^{2\pi i\frac{n_0}{\pi_0}},e^{i\theta_0}), &
 h &= (e^{2\pi i\frac{n_1}{\pi_1}},e^{i\theta_1},e^{i\tilde\theta_1}), & & \\
 l &= (e^{2\pi i\frac{n_2}{\pi_2}},e^{2\pi i\frac{\tilde n_2}{\tilde\pi_2}},e^{i\theta_2}), &
 m &= (e^{2\pi i\frac{n_3}{\pi_3}},e^{i\theta_3}), &
 n &= e^{2\pi i\frac{n_4}{\pi_4}},
\end{align}
the maps are given as follows:
\begin{itemize}
  \item Boundary maps:
  \begin{align}
    \partial_1 h &= (1,e^{2\pi i\frac{n_1}{\pi_1}P_1}) = e_G &
    \partial_2 l &= (1,e^{2\pi in_2\frac{P_2}{\pi_2}},e^{2\pi in_2\frac{\tilde P_2}{\tilde\pi_2}}) \\
    \partial_3 m &= (1,1,e^{2\pi in_3\frac{P_3}{\pi_3}}) &
    \partial_4 n &= (1,e^{2\pi in_4\frac{P_4}{\pi_4}})
  \end{align}
  \item Left actions:
  \begin{align}
 \act{g}h &= (e^{2\pi i\frac{n_1}{\pi_1}},e^{i\theta_1 + 2\pi i\frac{p_{2,2}P_2/p_2}{\pi_0\pi_1}n_0n_1},e^{i\tilde\theta_1 - 2\pi i\frac{p_{2,1}\tilde P_2/p_2}{\pi_0\pi_1}n_0n_1}) \\
 \act{g}l &= (e^{2\pi i\frac{n_2}{\pi_2}},e^{2\pi i\frac{\tilde n_2}{\tilde\pi_2}},e^{i\theta_2 + 2\pi i\frac{p_{2,1}}{\pi_2\pi_0}n_0n_2 + 2\pi i\frac{p_{2,2}}{\tilde \pi_2\pi_0}\tilde n_2n_0})
  \end{align}
  \item Liftings:
  \begin{align}
 \{h,h'\} &= (1,1,e^{2\pi i\frac{3p_1}{\pi_1^2}n_1n_1'}) \\
 \{h,l\}' &= \left(1,e^{2\pi i\frac{p_{2,1}}{\pi_2\pi_1}n_2n_1 + 2\pi i\frac{p_{2,2}}{\tilde\pi_2\pi_1}\tilde n_2n_1}\right)
  \end{align}
\end{itemize}
The boundary maps are a characteristic feature of TAE. They map electric symmetries to magnetic symmetries. We also have actions of the 0-form symmetry on the 2-form and 3-form symmetries. On the other hand, there are no nontrivial actions of the 1-form and 2-form symmetries on higher-form symmetries, since their codimensions are too large to allow linking with other topological operators. Among the liftings, the $HH$- and right $HL$-Peiffer liftings are nontrivial. The $HH$-Peiffer lifting can be interpreted as an induced topological defect of the 2-form magnetic symmetry localized at the intersection of two 1-form symmetry operators. The most notable feature is the presence of a nontrivial right $HL$-Peiffer lifting because the term $\scA_{(2)}\HLawedge\scA_{(3)}+\scA_{(2)}\HLbwedge\scA_{(3)}$ has non-zero value. This is one of the characteristic operations of the 3-crossed module. Here, we define the left $HL$-Peiffer lifting to be trivial, as it determines the action of $H$ on $L$, while the right $HL$-Peiffer lifting determines the action of $\partial H\subset G$ on $L$. 
\begin{equation}
 \{h,l\} = 1_M
\end{equation}
The liftings called Homanian and $LL$-Peiffer are trivial in the present case, as they do not appear in the fake curvatures. 
\begin{align}
 \{l,l'\} = \{h_z,h_y,h_x\} = \{h_z,h_y,h_x\}' = 1_M
\end{align}

With all operations now defined, one can verify that they satisfy the axioms of a 3-crossed module.

\subsection{Revisiting Gauge Transformations\label{sec:GTofBF}}
Finally, we turn to the gauge transformation laws, which we had postponed. As discussed in Sec.~\ref{sec:SRGT}, these transformation laws can be determined by requiring the invariance of the \Stueckelberg couplings. Alternatively, they can be derived by regarding the background fields as the gauge fields of a four-group gauge theory. As discussed in Sec.~\ref{sec:BriefIntro} and App.~\ref{App:4GT}, the theory admits four types of gauge transformations, referred to as the first kind through the fourth kind.\footnote{In addition, in the present case, there exists a fifth kind gauge transformation associated with the fifth group $N$.} We can also determine the gauge transformation laws for the dynamical gauge field $a_{(i)}$ by requiring that the corresponding \Stueckelberg coupling remain invariant. 

In what follows, we examine a gauge transformation law of the third kind. \footnote{The remaining gauge transformations are given in App.~\ref{App:derGT}.}
Let us denote the gauge transformation parameters collectively as $\lambda = (\Lambda_{(2)},\tilde\Lambda_{(2)},\lambda_{(2)})$. The gauge transformation of the third kind is given by:
\begin{align}
 \scA_{(2)}' &= \scA_{(2)} + \delta_2\lambda, \\
 \scA_{(3)}' &= \scA_{(3)} + \diff\lambda + \scA_{(1)}\actwedge\lambda, \\
 \scA_{(4)}' &= \scA_{(4)} - \scA_{(2)}'\HLbwedge\lambda,
\end{align}
where the remaining fields are invariant under this transformation. Since the relevant operations were defined in Sec.~\ref{sec:BFFGT}, these transformations can be written explicitly as:
\begin{align}
 \scA_{(2)}' &= (\A1^{\pi_0},\B2^{P_2} + P_2\Lambda_{(2)},\B2^{\tilde P_2} + P_2\tilde \Lambda_{(2)}) \\
 \scA_{(3)}' &= \left(\A3^{\pi_2} + \diff\Lambda_{(2)},\A3^{\tilde \pi_2} + \diff\tilde \Lambda_{(2)},\B3^{\tilde P_3} + \diff\lambda_{(2)} - \frac{p_{2,1}}{2\pi}A^{\pi_0}_{(1)}\Lambda_{(2)} - \frac{p_{2,2}}{2\pi}A^{\pi_0}_{(1)}\tilde\Lambda_{(2)}\right) \\
 \scA_{(4)}' &= \left(\A4^{\pi_3},\B4^{P_4} - \frac{p_{2,1}}{2\pi }A^{\pi_1}_{(2)}\Lambda_{(2)} - \frac{p_{2,2}}{2\pi}A^{\pi_1}_{(2)}\tilde\Lambda_{(2)}\right)
\end{align}
These transformations preserve the fake curvature invariant. Furthermore, by requiring the invariance of the \Stueckelberg coupling, we obtain the transformation laws for the dynamical gauge fields $a_{(i)}$ as well. In this case, only 2-form fields are transformed as:
\begin{align}
 a_{(2)} &\to a_{(2)} + \Lambda_{(2)}, & \tilde a_{(2)} &\to \tilde a_{(2)} + \tilde\Lambda_{(2)}.
\end{align}
These results precisely match those derived in Sec.~\ref{sec:SRGT}.

\section{Summary and Discussion\label{sec:summary-discussion}}

In this paper, we examined the higher-form symmetry structure that arises in the higher-dimen\-sional extension of the TAE, and showed that it can be described in terms of the 3-crossed module introduced in our previous work. The main goal was to verify that the structure extracted from physical considerations indeed captures the higher-form symmetry content of the theory.

In Sec. \ref{sec:action}, we extended the TAE to five dimensions. This was achieved by adding possible BF-like terms and topological interactions 
constructed from higher form gauge fields with appropriate degrees. In particular, we have introduced the topologically massive axion and photon, as well as higher-form gauge fields that interact topologically with the axion and photon. In Sec. \ref{sec:GS}, we constructed electric and magnetic symmetry operators from equations of motion and Bianchi identities, respectively. We found that, for the electric symmetries, the transformation parameters are constrained by gauge invariance. 

In Sec. \ref{sec:BGgauge}, we performed background gauging for electric symmetries via the \Stueckelberg coupling to understand them in more detail. We worked on the six-dimensional auxiliary manifold whose boundary is the spacetime $M_5$, where gauge variance is translated into the dependence on the choice of the auxiliary manifold. This dependence is described by a topological action on the bulk. However, a naive \Stueckelberg coupling fails due to the operator-valued ambiguities. At first glance, these ambiguities resemble 't Hooft anomalies, but as their name suggests, they take values in $q$-numbers. They are not true anomalies, since they can be cancelled by appropriately modifying the couplings and gauge transformation laws, leading to a consistent theory \cite{Hidaka:2020iaz,Hidaka:2020izy,Hidaka:2021kkf,Hidaka:2021mml,Benini:2018reh}. These modifications encode the higher-group structure in the system. After this procedure, topological bulk terms remain, which are related to the anomaly inflow action associated with the 't Hooft anomalies. In addition, we also performed background gauging for the magnetic symmetries at the end of this section. 

In Sec. \ref{sec:str}, we reformulated these symmetries in terms of the four-group gauge theory, and extracted a 3-crossed module structure. Similar to the TAE case \cite{Hidaka:2021kkf,Hidaka:2021mml}, the modifications of the background gauge fields and the flat connection conditions can be interpreted as vanishing fake curvature conditions. To make this interpretation clearer, we first combined background gauge fields for each degree and then defined how the 3-crossed module operations act on these fields. From these operations, we could identify a 3-crossed module structure, since background fields encode the higher-group structure of the theory, as discussed above. In particular, we found that the $HL$-Peiffer lifting, one of the 3-crossed module operations, is nontrivial, and these operations satisfy the axioms of a 3-crossed module given in App.~\ref{App:3CM}. Finally, we showed that the modified gauge transformation laws introduced in Sec. \ref{sec:BGgauge} can be systematically obtained from the general transformation laws of four-group gauge theory. 

Our analysis confirms that the 3-crossed module, motivated by higher-group symmetry data, consistently reproduces the higher-group symmetry structure of the theory. In particular, we identified a nontrivial $HL$-Peiffer lifting, which constitutes one of the key operations of the 3-crossed module. We believe that the present work provides a concrete demonstration of how partial algebraic data, such as 3-crossed modules, can be extracted, at least partially, from physical symmetry structures.

Although our model does not exhibit Homanians, i.e., nontrivial triple brackets $\{-,-,-\}: H\times H\times H\to M$, and we did not analyze a naive six-dimensional extension of the TAE explicitly in this paper, a similar structure appears there, such as the triple intersection $\left<U_1U_1U_1\right> \sim \left<V_4\right>$, where $U_1$ and $V_4$ denote the 1-form and 4-form symmetry operators, respectively, as discussed in Ref.~\cite{Nakajima:2022feg}. This observation suggests that richer algebraic structures, potentially described by 4-crossed modules, may emerge in higher-dimensional theories. A detailed investigation of this possibility is left for future work.

\section*{Acknowledgements}
The work of MF is supported by JST SPRING, Grant No.~JPMJSP2114.
This work of RY is supported 
by JSPS KAKENHI Grant No.~JP25K17394.

\appendix
\section{Crossed Modules\label{CM}}
In this appendix, we briefly summarize the axioms of a 3-crossed module. To define this, we first recall the axioms of 1- and 2-crossed modules. The motivation for introducing the Peiffer lifting provides a key insight into how the axioms of a 2-crossed module extend to those of a 3-crossed module. For a more detailed discussion, we refer the reader to our previous work \cite{Fukuda2025}. We conclude this appendix with a concrete example of a 3-crossed module. For applications of crossed modules in physics, see, for example, Ref. \cite{Kapustin:2013uxa,Hidaka:2020iaz,Hidaka:2020izy,Hidaka:2021kkf,Hidaka:2021mml,Girelli:2007tt,Radenkovic:2019qme}. 

\subsection{(1-)Crossed Module}
A crossed module consists of a pair of groups together with structure maps satisfying the conditions below. We denote such a structure by $(H\overset{\partial}{\to}G,\triangleright)$.
\begin{itemize}
  \item $H$ and $G$ are groups. It is often convenient to write $G_1 = H$ and $G_0 = G$, where the subscripts indicate the degrees. 
  \item A map $\partial:~H\to G$ is a $G$ group homomorphism, called boundary map. Unless there is confusion, we often omit parentheses and write $\partial(h)$ as $\partial h,\ h\in H$ whenever no confusion arises. 
  \item $\triangleright$ denotes a left actions of $G$ by a automorphism on $H$ (together with the adjoint action on $G$ itself). We use the following shorthand:
  \begin{align}
    \label{eq:1CMGact}
 \act{g}h &:= g\triangleright h, & \act{g}g' &:= g\triangleright g' = gg'g^{-1}.
  \end{align}
 for $g,g'\in G$ and $h\in H$. 
  \item The operation $\partial$ is $G$ invariant, i.e.
  \begin{equation}
    \label{eq:1CMGinv}
 \act{g}\partial h = \partial \act{g}h.
  \end{equation}
  \item The action satisfies the Peiffer identity:
  \begin{equation}
    \label{eq:1CMPeifferid}
 \act{\partial h}h' = hh'h^{-1}
  \end{equation}
\end{itemize}
A tuple $(H\overset{\partial}{\to}G,\triangleright)$ satisfying the above conditions is called a (1-)crossed module.

These algebraic relations admit a convenient interpretation in terms of diagrams. We represent elements of $G$ by segments and elements of $H$ by squares, which can be understood as link variables and plaquettes in lattice gauge theories, respectively: 
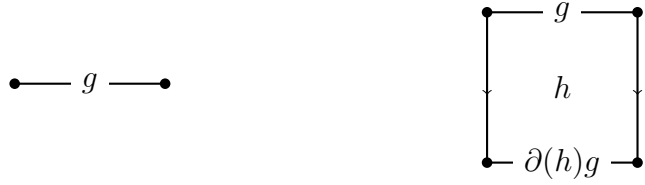
\begin{center}
  \hfill
  \begin{tikzpicture}[baseline=(T.base)]
    \draw [thick] (-1,0) to node (T) [fill=white] {$g$} (1,0);
    \fill (-1,0) circle [radius=2pt];
    \fill (1,0) circle [radius=2pt];
  \end{tikzpicture}
  \hfill
  \begin{tikzpicture}[baseline=(T.base)]
    \node (T) at (0,0) {$h$};
    \fill (1,1) circle [radius=2pt] coordinate (A);
    \fill (1,-1) circle [radius=2pt] coordinate (B);
    \fill (-1,1) circle [radius=2pt] coordinate (C);
    \fill (-1,-1) circle [radius=2pt] coordinate (D);
    \draw [thick] (A) to node [fill=white] {$g$} (C);
    \draw [thick] (B) to node [fill=white] {$\partial (h)g$} (D);
    \draw [thick] (A) to node [sloped] {\tikz[baseline=(T.base)]{\node (T) at (0,0) {};\draw [->] (0,0)to(0.1,0);}} (B);
    \draw [thick] (C) to node [sloped] {\tikz[baseline=(T.base)]{\node (T) at (0,0) {};\draw [->] (0,0)to(0.1,0);}} (D);
  \end{tikzpicture}
  \hfill\ \ 
\end{center}
In this picture, an element $h\in H$ act as a map $G\to G$ via the boundary homomorphism:
\begin{equation}
 h:~g\mapsto (\partial h)g
\end{equation}
The upper and lower edges of the square correspond to the source and target elements of $G$, respectively, indicated by the arrows on the sides (these arrows will be omitted when the context is clear). Multiplication in $G$ is represented by the concatenation of segments, whereas multiplication in $H$ corresponds to vertical stacking of squares.
\begin{align}
  \begin{tikzpicture}[baseline=-2pt]
    \draw [thick] (-1,0) to node (T) [fill=white] {$gg'$} (1,0);
    \fill (-1,0) circle [radius=2pt];
    \fill (1,0) circle [radius=2pt];
  \end{tikzpicture}
 \quad 
  &=
 \quad
  \begin{tikzpicture}[baseline=-2pt]
    \draw [thick] (-2,0) to node [fill=white] {$g$} (0,0) to node [fill=white] {$g'$} (2,0);
    \fill (-2,0) circle [radius=2pt];
    \fill (0,0) circle [radius=2pt];
    \fill (2,0) circle [radius=2pt];
  \end{tikzpicture} \\
  \begin{tikzpicture}[baseline=-2pt]
    \node (T) at (0,0) {$hh'$};
    \draw (-1/2,-1/2) rectangle (1/2,1/2);
  \end{tikzpicture}
 \qquad
  &= 
 \qquad
  \begin{tikzpicture}[baseline=-2pt]
    \node (T) at (0,0) {};
    \node at (0,-1/2) {$h$};
    \node at (0,1/2) {$h'$};
    \draw (-1/2,-1) rectangle (1/2,0);
    \draw (-1/2,0) rectangle (1/2,1);
  \end{tikzpicture}
\end{align}

The relations~\eqref{eq:1CMGact} and \eqref{eq:1CMGinv} can also be expressed diagrammatically. 
\begin{align}
  \begin{tikzpicture}[baseline=-2pt]
    \fill (-2,1) circle [radius=2pt] coordinate (A);
    \fill (0,1) circle [radius=2pt] coordinate (B);
    \fill (2,1) circle [radius=2pt] coordinate (C);
    \fill (-2,-1) circle [radius=2pt] coordinate (D);
    \fill (0,-1) circle [radius=2pt] coordinate (E);
    \fill (2,-1) circle [radius=2pt] coordinate (F);
    \node at (-1,0) {$h_1$};
    \node at (1,0) {$h_2$};
    \draw [thick] (A) to node [fill=white] {$g_1$} (B) to node [fill=white] {$g_2$} (C);
    \draw [thick] (D) to node [fill=white] {$g_1'$} (E) to node [fill=white] {$g_2'$} (F);
    \draw [thick] (A) to node [sloped] {\tikz[baseline=(T.base)]{\node (T) at (0,0) {};\draw [->] (0,0)to(0.1,0);}} (D);
    \draw [thick] (B) to node [sloped] {\tikz[baseline=(T.base)]{\node (T) at (0,0) {};\draw [->] (0,0)to(0.1,0);}} (E);
    \draw [thick] (C) to node [sloped] {\tikz[baseline=(T.base)]{\node (T) at (0,0) {};\draw [->] (0,0)to(0.1,0);}} (F);
  \end{tikzpicture}
 = 
  \begin{tikzpicture}[baseline=-2pt]
    \node (T) at (0,0) {$h_1\act{g_1}h_2$};
    \fill (1,1) circle [radius=2pt] coordinate (A);
    \fill (1,-1) circle [radius=2pt] coordinate (B);
    \fill (-1,1) circle [radius=2pt] coordinate (C);
    \fill (-1,-1) circle [radius=2pt] coordinate (D);
    \draw [thick] (A) to node [fill=white] {$g_1g_2$} (C);
    \draw [thick] (B) to node [fill=white] {$g_1'g_2'$} (D);
    \draw [thick] (A) to node [sloped] {\tikz[baseline=(T.base)]{\node (T) at (0,0) {};\draw [->] (0,0)to(0.1,0);}} (B);
    \draw [thick] (C) to node [sloped] {\tikz[baseline=(T.base)]{\node (T) at (0,0) {};\draw [->] (0,0)to(0.1,0);}} (D);
  \end{tikzpicture}
\end{align}
This figure states that the relation
\begin{equation}
  \partial(h_1\act{g_1}h_2)g_1g_2 = g_1'g_2'
\end{equation}
is satisfied, where we set $g_1' = \partial (h_1)g_1$ and $g_2' = \partial (h_2)g_2$. Thus, $G$ action on $H$ encodes how squares may be composed horizontally, in other words, how the multiplication of $G$ interacts with elements of $H$.

There are, however, two distinct ways to compose two squares horizontally:
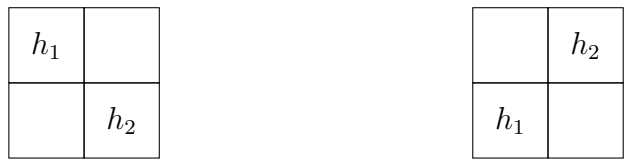
\begin{center}
  \hfill
  \begin{tikzpicture}[baseline=-2pt]
    \node (T) at (0,0) {};
    \node at (-1/2,1/2) {$h_1$};
    \node at (1/2,-1/2) {$h_2$};
    \draw (-1,-1) rectangle (0,0);
    \draw (-1,0) rectangle (0,1);
    \draw (0,-1) rectangle (1,0);
    \draw (0,0) rectangle (1,1);
  \end{tikzpicture}
  \hfill
  \begin{tikzpicture}[baseline=-2pt]
    \node (T) at (0,0) {};
    \node at (-1/2,-1/2) {$h_1$};
    \node at (1/2,1/2) {$h_2$};
    \draw (-1,-1) rectangle (0,0);
    \draw (-1,0) rectangle (0,1);
    \draw (0,-1) rectangle (1,0);
    \draw (0,0) rectangle (1,1);
  \end{tikzpicture}
  \hfill\ \ 
\end{center}
Here, identity elements are implicitly assigned to the empty regions. The diagram emphasizes that two composites differ a priori, depending on which mapping ($h_1$ or $h_2$) is applied first. The Peiffer identity \eqref{eq:1CMPeifferid} ensures that these two ways of composing squares agree. 

Indeed, the left-hand diagram corresponds to
\begin{equation}
  \label{eq:Left}
  \partial (\act{\partial h_1g_1}h_2h_1)g_1g_2
 = \partial (h_1\act{g_1}h_2)g_1g_2
 = (\partial h_1)g_1(\partial h_2)g_1^{-1}g_1g_2
 = (\partial h_1)g_1(\partial h_2)g_2 = g_1'g_2'
\end{equation}
while the right-hand diagram represents
\begin{equation}
  \label{eq:Right}
  \partial (h_1\act{g_1}h_2)g_1g_2 = \partial (h_1\act{g_1}h_2)g_1g_2 = (\partial h_1)g_1(\partial h_2)g_1^{-1}g_1g_2 = (\partial h_1)g_1(\partial h_2)g_2 = g_1'g_2'.
\end{equation}
From a categorical perspective, this is precisely the interchange law relating vertical and horizontal compositions in a strict 2-category.

\subsection{2-Crossed Module}
In this section, we present the axioms of a 2-crossed module. Motivated by categorical considerations and the extension to the 3-crossed module, we adopt a definition similar to that given in Ref.\cite{zbMATH07940412}. 

In the case of a 2-crossed module, an additional group $L$ enters the sequence of groups
\begin{equation}
 L\overset{\partial_2}{\to}H\overset{\partial_1}{\to}G.
\end{equation}
Here, the map $\partial_2:~L\to H$ is a group homomorphism satisfying
\begin{equation}
  \partial_1\circ\partial_2(l) = e_G,\qquad l\in L,
\end{equation}
so that elemetns of $L$ map into $\ker(\partial_1)$. This allows us to introduce \textit{twists} to the relations appearing in a crossed module. In this case, an additional operation called the Peiffer lifting is introduced. 

From the Eqs.~\eqref{eq:Left} and \eqref{eq:Right}, we obtain
\begin{equation}
  \label{eq:comP}
  \partial_1(h_1\act{g_1}h_2) = \partial_1(\act{\partial h_1g_1}h_2h_1).
\end{equation}
which expressed the compatibility condition between the two directions of multiplication associated with $G$ and $H$. As we discussed above, in the crossed module, this condition is satisfied because of the Peiffer identity \eqref{eq:1CMPeifferid}. However, the equation still holds as long as the expressions inside $\partial_1$ agree up to elements of $\ker(\partial_1)$. 

Since $\partial_1\circ\partial_2(l) = e_G$ for every $l\in L$, the consistency condition \eqref{eq:comP} remains valid if there exists an element of $L$, depending on $h_x = h_1$ and $h_y = \act{g_1}h_2$, that satisfies
\begin{equation}
  \partial_2\{h_x,h_y\} = h_xh_yh_x^{-1}\act{\partial h_x}h_y^{-1}.
\end{equation}
The map $\{\text{-},\text{-}\}:~H\times H\to L$ defined in this way is called Peiffer lifting, and can be interpreted as a twisted version of Peiffer identity \eqref{eq:1CMPeifferid}. Equivalently, it measures the nontriviality of the action of $\partial_1 H$ on $H$. Motivated by the above discussion, we define a 2-crossed module by the following axioms.

A 2-crossed module consists of the data
\begin{equation}
 (L\overset{\partial_2}{\to}H\overset{\partial_1}{\to}G,\triangleright,\{\text{-},\text{-}\})
\end{equation}
where:
\begin{itemize}
  \item $L,H$ and $G$ are groups. It is often convenient to write $G_0 = G,\ G_1 = H$ and $G_2 = L$, where the subscripts indicate the degrees.
  \item The maps $\partial_i,\ i=1,2$ are $G$ equivalent group homomorphisms. Unless there is confusion, we often omit the subscripts.
  \item $\triangleright$ denote left actions of lower degree group $G_i$ on higher degree group $G_j$ (for $j>i$) by automorphisms, with the action on $G_i$ itself given by conjugation.
  \item A $G$ equivalent map called the Peiffer lifting $\{\text{-},\text{-}\}:~H\times H\to L$, satisfying:
  \begin{itemize}
    \item [1.] Left actions:\footnote{Most literature, e.g., \cite{Hidaka:2020izy,Martins:2009evc,Wang:2013dwa}, defines $\{\partial l,\partial l'\} = ll'l^{-1}l'^{-1}$ as an axiom and derives $\act{h}l = l\{\partial l^{-1},h\}$ as a theorem. However, we take the latter as an axiom to emphasize the role of the action. This choice is more compatible with the categorical construction of 2-crossed modules. In our framework, the relation $\{\partial l,\partial l'\} = ll'l^{-1}l'^{-1}$ becomes a theorem, arising from the compatibility between the Peiffer-identity action and the $H$-action on $L$.}
    \begin{align}
      \partial\{h,h'\}\act{\partial h}h' &= hh'h^{-1}, &
 \act{h}l &= l\{\partial l^{-1},h\}, &
 \act{\partial h}l &= \act{h}l\{h,\partial l^{-1}\}, &
 \act{\partial l}l' &= ll'l^{-1}
    \end{align}
    \item [2.] Peiffer lifting with product inside them:
    \begin{align}
 \{h_zh_y,h_x\} &= \act{h_z}\{h_y,h_x\}\{h_z,\act{\partial h_y}h_x\}, &
 \{h_z,h_yh_x\} &= \{h_z,h_y\}\act{\act{\partial h_z}h_y}\{h_z,h_x\}
    \end{align}
 These encode the compatibility conditions for the action of $\partial_1H$ on $H$. 
  \end{itemize}
\end{itemize}

Therefore, when we enlarge an $n$-crossed module
\begin{equation}
 G_n\xrightarrow{\partial_n}\ldots\xrightarrow{\partial_1}G_0,
\end{equation}
by introducing a higher group $G_{n+1}$ on the left, it becomes possible to introduce additional twists in the $n$-crossed module relations. The allowed twists are determined by mathematical and physical considerations, such as those appearing in this paper, in Dijkgraaf-Witten theory, and in categorical constructions. Concretely, in the case of a 3-crossed module, one introduces twisted versions of the relations listed in items 1 and 2 above. 

\subsection{3-Crossed Module}
\label{App:3CM}
Continuing the same line of reasoning, one can derive the necessary structure of a 3-crossed module from the axioms of a 2-crossed module. Since the derivation is somewhat involved, we only present the resulting axioms of a 3-crossed module here. 

A \textit{3-crossed module} consists of the data
\begin{equation}
  (M\overset{\partial_3}{\to}L\overset{\partial_2}{\to}H\overset{\partial_1}{\to}G,\triangleright,\{\text{-},\text{-},\cdots\}),
\end{equation}
where
\begin{itemize}
  \item $M,L,H$ and $G$ are groups. It is often convenient to write $G_0 = G,\ G_1 = H,\ G_2 = L$ and $G_3 = M$ where subscripts indicate the degrees. We denote this by $\deg(G_i) = i$.
  \item The maps $\partial_i,\ i=1,2,3$ are $G$ equivalent group homomorphisms. Unless there is confusion, we often omit the subscripts.
  \item $\triangleright$ denotes left actions of lower degree group $G_i$ on higher degree group $G_j$ for $j>i$ by automorphisms, with the action on $G_i$ itself given by conjugation.
  \item $\{\text{-},\text{-},\cdots\}$ denotes six types of liftings listed in Table~\ref{tab:liftings}. The subscripts indicate the source of the corresponding maps, but unless there is confusion, we often omit them. 
  \begin{table}[ht]
    \begin{center}
      \begin{tabular}{c|c}
        Name & source and target \\\hline
        $HH$-Peiffer lifting &  $\{\text{-},\text{-}\}_{\text{\tiny HH}}~:~H\times H\to L$ \\ \hline
        $LL$-Peiffer lifting & $\{\text{-},\text{-}\}_{\text{\tiny LL}}~:~L\times L\to M$ \\ \hline
        \textit{left $HL$-Peiffer lifting} & $\{\text{-},\text{-}\}_{\text{\tiny HL}}~:~H\times L\to M$ \\ \hline
        \textit{right $HL$-Peiffer lifting} & $\{\text{-},\text{-}\}_{\text{\tiny HL}}'~:~H\times L\to M$ \\ \hline
        \textit{left Homanian} & $\{\text{-},\text{-},\text{-}\}_{\text{\tiny HHH}}~:~H\times H\times H\to M$ \\ \hline
        \textit{right Homanian} & $\{\text{-},\text{-},\text{-}\}_{\text{\tiny HHH}}'~:~H\times H\times H\to M$
      \end{tabular}
      \caption{Liftings}\label{tab:liftings}
    \end{center}
  \end{table}
\end{itemize}
These operations are required to satisfy the following axioms. In the axioms, the elements of a group $G$ are denoted by lowercase letters, e.g., $g,g',g_x,\ldots$.
\begin{itemize}
  \item [1.] Left actions:
  \begin{align}
    \partial\{h,h'\}\act{\partial h}h' &= hh'h^{-1} \\
    \partial\{h,l\}\act{h}l &= l\{\partial l^{-1},h\} & \partial\{h,l\}'\act{\partial h}l &= \act{h}l\{h,\partial l^{-1}\} \\
    \act{h}m &= m\{h,\partial m^{-1}\} & \act{\partial h}m &= \act{h}m\{h,\partial m^{-1}\}' \\
    \partial\{l,l'\}\act{\partial l}l' &= ll'l^{-1} \\
    \act{l}m &= m\{\partial m^{-1},l\} & \act{\partial l}m &= \act{l}m\{l,\partial m^{-1}\} \\
    \act{\partial m}m' &= mm'm^{-1}
  \end{align}
  \item [2.] Equivalence:
  \begin{align}
    \act{x}\partial y &= \partial\act{x}y \\
    \act{x}\{y,z,\ldots\} &= \{\act{x}y,\act{x}z,\ldots\}
  \end{align}
  where $\text{deg}(X) < \min(\text{deg}(Y),\deg(Z),\ldots)$. 
  \item [3.] Peiffer liftings with product inside them:
  \begin{align}
    \{h_zh_y,h_x\} &= \partial\{h_z,h_y,h_x\}\act{h_z}\{h_y,h_x\}\{h_z,\act{\partial h_y}h_x\} \\
    \{h_z,h_yh_x\} &= \partial\{h_z,h_y,h_x\}'\{h_z,h_y\}\act{\act{\partial h_z}h_y}\{h_z,h_x\} \\
    \{hh',l\} &= \act{l}\{\partial l^{-1},h,h'\}'\{h,l\}\act{h}\{h',l\} \\
    \{h,ll'\} &= \act{ll'}\{\partial l'^{-1},\partial l^{-1},h\}\act{ll'}\{l'^{-1},\{\partial l^{-1},h\}\}^{-1}\{h,l\}\act{\act{h}l}\{h,l'\} \\
    \{hh',l\}' &= \act{\act{hh'}l}\{h,h',\partial l^{-1}\}\act{h}\{h',l\}'\{h,\act{\partial h'}l\}' \\
    \{h,ll'\}' &= \act{\act{h}(ll')}\{h,\partial l'^{-1},\partial l^{-1}\}'\act{\act{h}l}\{h,l'\}'\act{\act{h}l\act{\partial h}l'}\{\act{\partial h}l'^{-1},\{h,\partial l^{-1}\}\}^{-1}\{h,l\}'
  \end{align}
  \item [4.] Homanian with product inside them:
  \begin{align}
    \label{eq:DHR}
    &\{h_wh_z,h_y,h_x\}\act{\act{h_wh_z}\{h_y,h_x\}}\{h_w,h_z,\act{\partial h_y}h_x\} = \{h_w,h_zh_y,h_x\}\act{h_w}\{h_z,h_y,h_x\} \\
    \label{eq:DHL}
    &\{h_w,h_zh_y,h_x\}'\{h_w,h_z,h_y\}' = \{h_w,h_z,h_yh_x\}'\act{\{h_w,h_z\}}(\act{\act{\partial h_w}h_z}\{h_w,h_y,h_x\}') \\
    \nonumber
    &\{h_w,h_z,h_yh_x\}\act{\act{h_w}\{h_z,h_yh_x\}}\{h_w,\act{\partial h_z}h_y,\act{\partial h_z}h_x\}'\act{h_w}\{h_z,h_y,h_x\}' \\
    \nonumber
    &= \{h_wh_z,h_y,h_x\}'\act{\{h_wh_z,h_y\}}(\act{\act{\partial (h_wh_z)}h_y}\{h_w,h_z,h_x\})\{h_w,h_z,h_y\} \\
    \label{eq:DHLR}
    &\hspace{1em}\times
    \act{\act{h_w}\{h_z,h_y\}}\{\{h_w,\act{\partial h_z}h_y\},\act{\act{\partial(h_wh_z)}h_yh_w}\{h_z,h_x\}\}
  \end{align}
  \item [5.] Switching the twist of Yang-Baxter:
  \begin{align}
    \nonumber
    1_M &= \{\partial\{h_z,h_y\},\act{\partial h_z}h_yh_z,h_x\}\{\{h_z,h_y\},\{\act{\partial h_z}h_yh_z,h_x\}\}^{-1}\act{\{h_z,h_y\}}\{\act{\partial h_z}h_y,h_z,h_x\} \\
    \nonumber
    &\hspace{1.5em} \{h_z,h_y,h_x\}'^{-1}\act{\{h_z,h_yh_x\}}(\act{\act{\partial h_z}\{h_y,h_x\}}\{\act{\partial (h_zh_y)}h_x,\{h_z,h_y\}^{-1}\}) \\
    \nonumber
    &\hspace{1.5em} \{h_z,\partial\{h_y,h_x\},\act{\partial h_y}h_xh_y\}'\act{\act{h_z}\{h_y,h_x\}}\{h_z,\{h_y,h_x\}^{-1}\}' \\
    \nonumber
    &\hspace{1.5em} \act{\act{h_z}\{h_y,h_x\}\act{\partial h_z}\{h_y,h_x\}^{-1}}\{\act{\partial h_z}\{h_y,h_x\},\{h_z,\act{\partial h_y}h_xh_y\}\}^{-1} \\
    &\hspace{1.5em} \act{\act{h_z}\{h_y,h_x\}}\{h_z,\act{\partial h_y}h_x,h_y\}'\{h_z,h_y,h_x\}^{-1}
  \end{align}
  \item [6.] Cubic relation:
  \begin{align}
    \act{l}\{\partial l',l^{-1}\}\act{ll'l^{-1}l'^{-1}}\{l',l\} &= \act{\act{\partial l}l'}\{\partial l,l'^{-1}\}'\{l,l'\}^{-1}
  \end{align}
\end{itemize}

\subsection{Example}
Here, we present a concrete example of a 3-crossed module. Let $\bG_2 = (C\to B\to A,\tilde\triangleright,\{\text{-},\text{-}\})$ be a given 2-crossed module. We define the groups $G,H,L,M$ and operations of the 3-crossed module as follows:
\begin{itemize}
  \item Groups,
  \begin{align}
    G &= A, & H &= B\times A, & L &= C\times B, & M &= C,
  \end{align}
  \item Boundary maps:
  \begin{equation}
    \begin{tikzpicture}[baseline=(EG1.base)]
      \node (M) at (-3,0) {$M$};
      \node (L) at (-1,0) {$L$};
      \node (H) at (1,0) {$H$};
      \node (G) at (3,0) {$G$};
      \draw [->] (M.east) to node [anchor=south] {$\partial_3$} (L.west);
      \draw [->] (L.east) to node [anchor=south] {$\partial_2$} (H.west);
      \draw [->] (H.east) to node [anchor=south] {$\partial_1$} (G.west);
      \node (EM1) at (-3,-2.5em) {$c$};
      \node (EL1) at (-1,-2.5em) {$(c,e_B)$};
      \node (EL2) at (-1,-4em) {$(c,b)$};
      \node (EH1) at (1,-4em) {$(b,e_A)$};
      \node (EH2) at (1,-5.5em) {$(b,a)$};
      \node (EG1) at (3,-5.5em) {$a$};
      \draw [|->] (EM1.east) to (EL1.west);
      \draw [|->] (EL2.east) to (EH1.west);
      \draw [|->] (EH2.east) to (EG1.west);
    \end{tikzpicture},
  \end{equation}
  \item Actions:
  \begin{align}
    \label{eq:3CMExLA1}
    \act{g}h &= (\act{a}b',\act{a}a'), & \act{g}l &= (\act{a}c,\act{a}b), & \act{g}m &= \act{a}c, \\
    \label{eq:3CMExLA2}
    && \act{h}l &= (\act{b}c',\act{b}b'), & \act{h}m &= \act{b}c, \\
    \label{eq:3CMExLA3}
    &&&& \act{l}m &= \act{c}c'.
  \end{align}
  Here and in the formulas below, all actions and liftings appearing on the right-hand side are the ones coming from the 2-crossed module $\bG_2$. 
  \item Liftings:
  \begin{itemize}
    \item $HH$-Peiffer lifting: for $h = (b,a),\ h' = (b',a')\in H,$
    \begin{equation}
      \{h,h'\} = (\{b,b'\},\act{b}b'\act{a}b'^{-1}),
    \end{equation}
    \item Left- and right Homanians: for $h_z = (b_z,a_z),h_y = (b_y,a_y),h_x = (b_x,a_x)\in H,$
    \begin{align}
      \{h_x,h_y,h_x\} &= \{b_zb_y,b_x\}\{b_z,\act{a_y}b_x\}^{-1}\act{b_z}\{b_y,b_x\}^{-1}, \\ \{h_z,h_y,h_x\}' &= \{b_z,b_yb_x\}\act{\act{a_z}b_y}\{b_z,b_y\}^{-1}\{b_z,b_y\}^{-1},
    \end{align}
    \item left and right $HL$-Peiffer: for $h = (b,a) \in H,\ l = (c,b)\in L$,
    \begin{align}
      \{h,l\} &= c'\{b'^{-1},b\}\act{b}c'^{-1}, & \{h,l\}' &= \act{b}c'\{b,b'^{-1}\}\act{a}c'^{-1},
    \end{align}
    \item $LL$-Peiffer: for $l = (c,b),\ l' = (c',b')\in L,$
    \begin{equation}
      \{l,l'\} = \act{c}c'\act{b}c'^{-1}.
    \end{equation}
  \end{itemize}
\end{itemize}
One can verify that the above data satisfy all axioms of a 3-crossed module given above.

\section{4-Group Gauge Theory\label{App:FG}}
As in the ordinary group case, higher principal bundles for higher-groups can be constructed in terms of the collections of transition functions. One can also define notions of curvature and holonomy. For example, in the case of a crossed module (two-group), connections and fake curvature were introduced in \cite{Breen:2001ie}, and it was shown that the vanishing fake curvature compensates for the existence of surface holonomy (also called Wilson surfaces)\cite{Baez:2004in}. Building on these works, \cite{Schreiber:2007S} proposed that holonomy or parallel transport is the essence of gauge theory, which naturally allows for the extension to higher-dimensional transport \cite{Schreiber:2008S}. Following this line, analogous constructions exist for 2-crossed modules (three-groups), where fake curvature and holonomy have been defined, and their relations discussed in \cite{Martins:2009evc,Wang:2013dwa}. 

For our 3-crossed module, we similarly expect that one can construct curvature and holonomy and study their relations. In this appendix, we present the fake curvatures and analyze their gauge transformation laws. However, if one attempts to derive the explicit forms of the fake curvatures directly from the flatness condition of a principal 4-bundle, following the same procedure as in the 1- and 2-crossed module cases \cite{Baez:2004in,Saemann:2013pca}, the calculation becomes extremely cumbersome. Instead, we determine their forms by analogy with the case of three-group gauge theory. Concretely, this amounts to adding to the exterior derivatives of connections the possible wedge products deformed by the 3-crossed module operations.

\subsection{4-Lie Algebra}
For a group $G_i$ of degree $i$ in a higher-group, the associated connection is given by an $(i+1)$-form $\A{i+1}$ taking values in the Lie algebra $\mathfrak{g}_i$. To define such connections, we need the Lie algebra analogue of a 3-crossed module, which we call \textit{differential 3-crossed module}. In this section, we present the linearized axioms of a 3-crossed module. 

A differential 3-crossed module consists of the data
\begin{equation}
 (\mathfrak{m}\overset{\delta_3}{\to}\mathfrak{l}\overset{\delta_2}{\to}\mathfrak{h}\overset{\delta_1}{\to}\mathfrak{g},\triangleright,\{\text{-},\cdots\}),
\end{equation}
where
\begin{itemize}
  \item $\mathfrak{m},\mathfrak{l},\mathfrak{h}$ and $\mathfrak{g}$ are Lie algebras associated with the groups $M,L,H$ and $G$ of a 3-crossed module.
  \item The maps $\delta_i,\ i=1,2,3$ are linear maps. 
  \item $\triangleright$ denotes the action by derivations of a lower degree Lie algebra $\mathfrak{g}_i$ on a higher degree one $\mathfrak{g}_j$ for $j>i$, with the action on $\mathfrak{g}_i$ itself given by the adjoint representation. 
  \item $\{\text{-},\text{-},\cdots\}$ denotes the six types of multilinear maps listed in Table~\ref{tab:diffliftings}. As in the case of a 3-crossed module, we refer to these maps as liftings.
  \begin{table}[ht]
    \begin{center}
      \begin{tabular}{c|c}
 Name & source and target \\\hline
        $\mathfrak{hh}$-Peiffer lifting &  $\{\text{-},\text{-}\}_{\text{\tiny $\mathfrak{hh}$}}:~\mathfrak{h}\times\mathfrak{h} \to \mathfrak{l}$ \\ \hline
        $\mathfrak{ll}$-Peiffer lifting & $\{\text{-},\text{-}\}_{\text{\tiny $\mathfrak{ll}$}}:~\mathfrak{l}\times\mathfrak{l} \to \mathfrak{m}$ \\ \hline
        \textit{left $\mathfrak{hl}$-Peiffer lifting} & $\{\text{-},\text{-}\}_{\text{\tiny $\mathfrak{hl}$}}:~\mathfrak{h}\times\mathfrak{l} \to \mathfrak{m}$ \\ \hline
        \textit{right $\mathfrak{hl}$-Peiffer lifting} & $\{\text{-},\text{-}\}_{\text{\tiny $\mathfrak{hl}$}}':~\mathfrak{h}\times\mathfrak{l} \to \mathfrak{m}$ \\ \hline
        \textit{left Homanian} & $\{\text{-},\text{-},\text{-}\}_{\text{\tiny $\mathfrak{hhh}$}}:~\mathfrak{h}\times\mathfrak{h}\times\mathfrak{h} \to \mathfrak{m}$ \\ \hline
        \textit{right Homanian} & $\{\text{-},\text{-},\text{-}\}_{\text{\tiny $\mathfrak{hhh}$}}':~\mathfrak{h}\times\mathfrak{h}\times\mathfrak{h} \to \mathfrak{m}$
      \end{tabular}
      \caption{Liftings}\label{tab:diffliftings}
    \end{center}
  \end{table}
\end{itemize}
These operations are required to satisfy the following axioms. In the axioms, the elements of Lie groups $\mathfrak{g,h,l}$ and $\mathfrak{m}$ are denoted by $X,Y,Z$ and $W$ (and their indexed variants), respectively.
\begin{itemize}
  \item [1.] Left actions:
  \begin{align}
    \delta\{Y,Y'\} &= [Y,Y'] - \act{\delta Y}Y' \\
    \label{eq:fund1}
    \delta\{Y,Z\} &= -\{\delta Z,Y\} - \act{Y}Z & \delta\{Y,Z\}' &= \act{Y}Z - \{Y,\delta Z\} - \act{\delta Y}Z \\
    \act{Y}W &= -\{Y,\delta W\} & \act{\partial Y}W &= \act{Y}W - \{Y,\delta W\}' \\
    \label{eq:fund2}
    \delta\{Z,Z'\} &= [Z,Z'] - \act{\delta Z}Z' \\
    \act{Z}W &= -\{\delta W,Z\} & \act{\delta Z}W &= \act{Z}W - \{Z,\delta W\} \\
    \act{\delta W}W' &= [W,W']
  \end{align}
  \item [2.] Equivalence:
  \begin{align}
    \act{v}\delta u &= \delta\act{v}u \\
    \act{v}\{u_1,u_2,\cdots,u_n\} &= \sum_{i=1}^n\{u_1,u_2,\cdots,\act{v}u_i,\cdots,u_n\}
  \end{align}
 where $\text{deg}(v) < \min(\{\text{deg}(u_i)\})$. 
  \item [3.] Peiffer liftings with products in the arguments:
  \begin{align}
 \{[Y_z,Y_y],Y_x\} &= \delta\{Y_z,Y_y,Y_x\} + \act{Y_{{\lbrack}z}}\{Y_{y\rbrack},Y_x\} + \{Y_{{\lbrack}z},\act{\delta Y_{y\rbrack}}Y_x\} \\
 \{Y_z,[Y_y,Y_x]\} &= \delta\{Y_z,Y_y,Y_x\}' + \act{Y_y}\{Y_z,Y_x\} - \act{Y_x}\{Y_z,Y_y\} \\
 \{[Y,Y'],Z\} &= -\{\delta Z,Y,Y'\}' + \act{Y}\{Y',Z\} - \act{Y'}\{Y,Z\} \\
 \nonumber
 \{Y,[Z,Z']\} &= -\{\delta Z,\delta Z',Y\} \\
    &\hspace{1.15em} -\{Z',\{\delta Z,Y\}\} + \{Z,\{\delta Z',Y\}\} + \act{Z}\{Y,Z'\} - \act{Z}\{Y,Z'\} \\
 \{[Y,Y'],Z\}' &= -\{Y,Y',\delta Z\} + \act{{\lbrack}Y}\{Y'{\rbrack},Z\}' + \{{\lbrack}Y,\act{\delta Y'\rbrack}Z\}' \\
 \nonumber
 \{Y,[Z,Z']\}' &= -\{Y,\delta Z,\delta Z'\}' \\
    &\hspace{1.15em} + \act{Z}\{Y,Z'\}' - \act{Z'}\{Y,Z\}' - \{Z',\{Y,\delta Z\}\} + \{Z,\{Y,\delta Z'\}\}
  \end{align}
  \item [4.] Homanian operations with products in the arguments:
  \begin{align}
    \sum_{(z,y,x)}(\{[Y_z,Y_y],Y_x,Y\} + \{Y_z,Y_y,\act{\delta Y_x}Y\} - \act{Y_x}\{Y_z,Y_y,Y\}) &= 0 \\
    \sum_{(z,y,x)}(\{Y,Y_z,[Y_y,Y_x]\}' + \act{Y_z}\{Y,Y_y,Y_x\}') &= 0
  \end{align}
  \vspace{-1em}
  \begin{align}
 \nonumber
    &\{Y_w,Y_z,[Y_y,Y_x]\} + \act{Y_{{\lbrack}w}}\{Y_{z\rbrack},Y_y,Y_x\}' + \{Y_{{\lbrack}w},\act{\delta Y_{z\rbrack}}Y_y,Y_x\}' + \{Y_{{\lbrack}w},Y_y,\act{\delta Y_{z\rbrack}}Y_x\}' \\
    &= \{[Y_w,Y_z],Y_y,Y_x\}' + \act{Y_{{\lbrack}y}}\{Y_w,Y_z,Y_{x\rbrack}\} + \{\{Y_w,Y_y\},\{Y_z,Y_x\}\} - \{\{Y_w,Y_x\},\{Y_z,Y_y\}\}
  \end{align}
  \item [5.] Compatibility relations for twisted Yang-Baxter:
  \begin{align}
 \{Y_z,\{Y_y,Y_x\}\}' + \{Y_x,\{Y_z,Y_y\}\} + \{Y_z,Y_y,Y_x\}' + \{Y_z,Y_y,Y_x\} &= 0
  \end{align}
  \item [6.] Cubic relation:
  \begin{align}
 \{\delta Z',Z\} - \{Z',Z\} &= \{\delta Z,Z'\}' + \{Z,Z'\}
  \end{align}
\end{itemize}

\subsection{Four-Group Gauge Transformation Laws}\label{App:4GT}
We now introduce the fake curvatures. As mentioned above, we expect that the fake curvatures can be expressed as the sum of the exterior derivatives of the gauge fields, $\diff\A{i+1}$, together with all possible terms constructed using the operations of a differential 3-crossed module:
\begin{align}
  \Omega_1 &= \diff \A1 + \frac{1}{2}\A1\liewedge \A1 - \delta\A2, \\
  \Omega_2 &= \diff \A2 + \A1\actwedge\A2 - \delta\A3, \\
  \Omega_3 &= \diff \A3 + \A1\actwedge\A3 + \A2\pwedge\A2 - \delta\A4, \\
  \Omega_4 &= \diff \A4 + \A1\actwedge\A4 + \A2\HLawedge\A3 + \A2\HLbwedge\A3.
\end{align}

Next, let us consider the gauge transformations that preserve the vanishing of the fake curvature $\Omega_i = 0$. Under these transformations, the fake curvatures transform covariantly. In addition to the ordinary gauge transformation, there can be shift transformations associated with the boundary maps $\delta_i$. If a gauge field $\A{i}$ is shifted by a gauge parameter $\Lambda_{(i)}$ valued in Lie algebra $\mathfrak{g}_i$ as $\A{i} + \delta_i\Lambda_{(i)}$, then the $j$-form gauge fields with $j>i$ must also be transformed to preserve the vanishing condition. After some tedious computations, we obtain the following four types of transformation laws.
\begin{itemize}
  \item 4-gauge transformation of the first kind:
  \begin{align}
    \label{eq:GT1A1}
 \A1\to \A1' &= g^{-1}\A1g + g^{-1}\diff g \\
 \A2\to \A2' &= \act{g^{-1}}\A2 \\
 \A3\to \A3' &= \act{g^{-1}}\A3 \\
    \label{eq:GT1A4}
 \A4\to \A4' &= \act{g^{-1}}\A4
  \end{align}
 where the left action of $G$ on higher-groups is defined via pushforward. Under this transformation, the fake curvatures transform as:
  \begin{align}
    \Omega_1 \to \Omega_1' &= g^{-1}\Omega_1g \\
    \Omega_i \to \Omega_i' &= \act{g^{-1}}\Omega_i,\qquad (i=3,4,5)
  \end{align}
  \item 4-gauge transformation of the second kind:
  \begin{align}
    \label{eq:GT2A1}
 \A1\to \A1' &= \A1 + \delta\eta \\
 \A2\to \A2' &= \A2 + \diff\eta + \A1\actwedge\eta + \frac{1}{2}\eta\liewedge\eta \\
 \A3\to \A3' &= \A3 - \A2\pwedge\eta - \eta\pwedge\A2' \\
    \label{eq:GT2A4}
 \A4 \to \A4' &= \A4 - \eta\HLawedge\A3 - \eta\HLbwedge\A3' + \frac{1}{2}\{\A2,\eta,\eta\}' + \frac{1}{2}\{\eta,\eta,\A2'\}
  \end{align}
 Under this transformation, the fake curvatures transform as:
  \begin{align}
    \Omega_1\to \Omega_1' &= \Omega_1 \\
    \Omega_2\to \Omega_2' &= \Omega_2 + \Omega_1\actwedge\eta \\
    \Omega_3\to \Omega_3' &= \Omega_3 - \Omega_2\pwedge\eta + \eta\pwedge\Omega_2' \\
    \Omega_4 \to \Omega_4' &= \Omega_4 + \eta\HLawedge\Omega_3 + \eta\HLbwedge\Omega_3' + \frac{1}{2}\{\Omega_2,\eta,\eta\}' + \frac{1}{2}\{\eta,\eta,\Omega_2'\}
  \end{align}
  \item 4-gauge transformation of the third kind:
  \begin{align}
    \label{eq:GT3A1}
 \A1 \to \A1' &= \A1 \\
 \A2 \to \A2' &= \A2 + \delta\lambda \\
 \A3 \to \A3' &= \A3 + \diff\lambda + \A1\actwedge\lambda \\
    \label{eq:GT3A4}
 \A4 \to \A4' &= \A4 - \A2\HLawedge\lambda - \A2'\HLbwedge\lambda - \lambda\pwedge\lambda
  \end{align}
 Under this transformation, the fake curvatures transform as:
  \begin{align}
    \Omega_1 \to \Omega_1' &= \Omega_1 \\
    \Omega_2 \to \Omega_2' &= \Omega_2 \\
    \Omega_3 \to \Omega_3' &= \Omega_3 + \Omega_1\actwedge\lambda \\
    \Omega_4 \to \Omega_4' &= \Omega_4 - \Omega_2\HLawedge\lambda - \Omega_2'\HLbwedge\lambda
  \end{align}
  \item 4-gauge transformation of the fourth kind:
  \begin{align}
    \label{eq:GT4A1}
 \A1\to \A1' &= \A1 \\
 \A2\to \A2' &= \A2 \\
 \A3\to \A3' &= \A3 + \delta\mu \\
    \label{eq:GT4A4}
 \A4\to \A4' &= \A4 + \diff\mu + \A1\actwedge\mu
  \end{align}
 Under this transformation, the fake curvatures transform as:
  \begin{align}
    \Omega_1\to \Omega_1' &= \Omega_1 \\
    \Omega_2\to \Omega_2' &= \Omega_2 \\
    \Omega_3\to \Omega_3' &= \Omega_3 \\
    \Omega_4\to \Omega_4' &= \Omega_4 + \Omega_1\actwedge\mu
  \end{align}
\end{itemize}

\subsection{Derivation of Gauge Transformations}\label{App:derGT}
In this appendix, we give the full set of gauge transformation laws which we omitted in Sec.~\ref{sec:SRGT} and \ref{sec:GTofBF}. 

\subsubsection{Second kind}
Let us define the gauge transformation parameter $\eta$ as follows:
\begin{equation}
  \eta = (\Lambda_{(1)},\lambda_{(1)},\tilde\lambda_{(1)})
\end{equation}
Using 3-crossed module operations on gauge fields \eqref{eq:deltaA2}--\eqref{eq:A2HLaA3}, and gauge transformation laws of second kind given in \eqref{eq:GT2A1}--\eqref{eq:GT2A4}, we obtain
\begin{align}
 \scA_{(2)}' &= \left(\A2^{\pi_1} + \diff\Lambda_{(1)},\B2^{P_2} + \diff\lambda_{(1)} + \frac{p_{2,2}P_2}{2\pi p_2}\Lambda_{(1)}A^{\pi_1}_{(2)},\B2^{\tilde P_2} + \diff\tilde\lambda_{(1)} - \frac{p_{2,1}\tilde P_2}{2\pi p_2}\Lambda_{(1)}A^{\pi_1}_{(2)}\right) \\
 \scA_{(3)}' &=  \left(\A3^{\pi_2},\A3^{\tilde \pi_2},\B3^{\tilde P_3} - \frac{3p_1}{\pi}A^{\pi_1}_{(2)}{}\Lambda_{(1)} - \frac{3p_1}{2\pi}\Lambda_{(1)}\diff\Lambda_{(1)}\right) \\
 \scA_{(4)}' &= \left(\A4^{\pi_3},\B4^{P_4} - \frac{p_{2,1}}{2\pi}A^{\pi_2}_{(3)}\Lambda_{(1)} - \frac{p_{2,2}}{2\pi}A^{\tilde \pi_2}_{(3)}\Lambda_{(1)}\right).
\end{align}
Gauge transformation laws for dynamical fields $a_{(i)}$ are determined by requiring the invariance of the \Stueckelberg couplings:
\begin{equation}
 a_{(1)} \to a_{(1)} + \Lambda_{(1)}.
\end{equation}
All other fields are invariant under this transformation.

\subsubsection{Fourth kind}
Let us define the gauge transformation parameter $\mu$ as follows:
\begin{equation}
  \mu = (\Lambda_{(3)},\lambda_{(3)})
\end{equation}
Using 3-crossed module operations on gauge fields \eqref{eq:deltaA2}--\eqref{eq:A2HLaA3}, and gauge transformation laws of fourth kind given in \eqref{eq:GT4A1}--\eqref{eq:GT4A4}, we obtain
\begin{align}
 \scA_{(4)}'
 = (\A4^{\pi_3} + \diff\Lambda_{(3)},\B4^{P_4} + \diff\lambda_{(3)}).
\end{align}
Gauge transformation laws for dynamical fields $a_{(i)}$ are determined by requiring the invariance of the \Stueckelberg couplings:
\begin{equation}
 a_{(3)} \to a_{(3)} + \Lambda_{(3)}.
\end{equation}
All other fields are invariant under this transformation.

\subsubsection{Fifth kind}
Let us define the gauge transformation parameter $\nu$ as follows:
\begin{equation}
  \nu = \Lambda_{(4)}.
\end{equation}
Since the only nontrivial 4-crossed module operation is the boundary map $\partial_4$ given in equation~\eqref{eq:deltaA5}, the gauge transformation for the field $\scA_{(5)}$ has only the exterior derivative term of the parameter:
\begin{align}
 \scA_{(4)}' &= (\A4^{\pi_3},\B4^{P_4} + P_4\Lambda_{(4)}), \\
 \scA_{(5)}' &= \A5^{\pi_4} + \diff\Lambda_{(4)}.
\end{align}
Gauge transformation laws for dynamical fields $a_{(i)}$ are determined by requiring the invariance of the \Stueckelberg couplings:
\begin{equation}
 a_{(4)} \to a_{(4)} + \Lambda_{(4)}.
\end{equation}
All other fields are invariant under this transformation.

\subsubsection{First kind}
We already have the action of the Lie algebra of the 0-form symmetry group on the Lie algebras of the higher symmetry groups. However, in the context of gauge transformations, we need the action of the \textbf{group} itself (not just its Lie algebra) on the Lie algebras of the higher-groups. 

A useful intuition can be obtained from the following example of a crossed module. Let $G = U(1) \ni g=e^{i\psi}$ and $H = U(1) \times U(1) \ni h=\mathrm{diag}(e^{i\theta_1},e^{i\theta_2})$, and define the boundary map and left action as follows:
\begin{align}
  \partial h &= e^{i\theta_1}, &
 \act{g}h &= \begin{pmatrix}e^{i\theta_1} & \\ & e^{i(\theta_2 + \psi\theta_1)}\end{pmatrix}.
\end{align}
With this setup, the complex $H \to G$ satisfies the axioms of a crossed module. The structure of the action is similar to the 3-crossed module that appears in our model. The action of a Lie group $G$ on the Lie algebra of $H$ can be defined via the pushforward. Let $X$ and $Y$ be Lie algebra elements corresponding to $g$ and $h$, respectively. The action is defined by:
\begin{equation}
 \act{g}Y := \left.\frac{\diff\ (g\triangleright c(t))}{\diff t}\right|_{t=0},
\end{equation}
where $c(t)$ is a curve in $H$ with $c(0) = e_H$ (the identity element of $H$) and $Y = \left.\diff c(t)/\diff t\right|_{t=0}$. Following this definition, we obtain:
\begin{equation}
 \act{g}Y = \begin{pmatrix}\theta_1 & \\ & \theta_2 + \psi\theta_1\end{pmatrix}
\end{equation}
Because the action of Lie algebra is $\act{X}Y = \mathrm{diag}(0,\psi\theta_1)$, the Lie group action $\act{g}Y$ is given by the sum of $Y$ and $\act{X}Y$.

Let us define $\gamma$ as follows:
\begin{equation}
  \gamma = (\Lambda_{(0)},\lambda_{(0)}).
\end{equation}
In the present case, the first kind gauge transformation is related to the gauge transformation for $(-1)$-form symmetry. Therefore, the 0-form gauge field transforms as
\begin{equation}
 \scA_{(0)}' = \scA_{(0)} + \delta_0\gamma = \B0 + P_0\Lambda_0.
\end{equation}
Now, there are two nontrivial actions of $G$ on the higher-groups $H$ \eqref{eq:A1actA2} and $L$ \eqref{eq:A1actA3}. Using these actions and gauge transformation laws of the first kind given in \eqref{eq:GT1A1}--\eqref{eq:GT1A4}, the first kind of gauge transformations is given by:
\begin{align}
 \scA_{(1)}' &= \scA_{(1)} - ig^{-1}\diff g
 = (\A1 + \diff\Lambda_{(0)},\B1 + \diff\lambda_{(0)}) \\
 \scA_{(2)}'
  &= \scA_{(2)} - \gamma\actwedge\scA_{(2)} 
 = (\A2^{\pi_1},\B2^{P_2} - \frac{p_{2,2}P_2}{2\pi p_2}\Lambda_{(0)}\A2^{\pi_1},\B2^{\tilde P_2} + \frac{p_{2,1}P_2}{2\pi p_2}\Lambda_{(0)}\A2^{\pi_1}) \\
 \scA_{(3)}'
  &= \scA_{(3)} - \gamma\actwedge\scA_{(3)} 
 = (\A3^{\pi_2},\A3^{\tilde \pi_2},\B3^{P_3} - \frac{p_{2,1}}{2\pi}\A3^{\pi_2}\Lambda_{(0)} - \frac{p_{2,2}}{2\pi}\A3^{\tilde \pi_2}\Lambda_{(0)})
\end{align}
Gauge transformation laws for dynamical gauge fields are determined in the same way as before:
\begin{equation}
 a_{(0)} \to a_{(0)} + \gamma^A_{(0)}
\end{equation}
All other fields are invariant under this transformation.


\bibliographystyle{ytphys}
\bibliography{ref.bib}

@article{Gaiotto:2014kfa,
    author = "Gaiotto, Davide and Kapustin, Anton and Seiberg, Nathan and Willett, Brian",
    title = "{Generalized Global Symmetries}",
    eprint = "1412.5148",
    archivePrefix = "arXiv",
    primaryClass = "hep-th",
    doi = "10.1007/JHEP02(2015)172",
    journal = "JHEP",
    volume = "02",
    pages = "172",
    year = "2015"
}

@article{Nakajima:2022feg,
    author = "Nakajima, Tatsuki and Sakai, Tadakatsu and Yokokura, Ryo",
    title = "{Higher-group structure in 2n-dimensional axion-electrodynamics}",
    eprint = "2211.13861",
    archivePrefix = "arXiv",
    primaryClass = "hep-th",
    reportNumber = "KEK-TH-2475",
    doi = "10.1007/JHEP01(2023)150",
    journal = "JHEP",
    volume = "01",
    pages = "150",
    year = "2023"
}

@article{Nakajima:2024vgc,
    author = "Nakajima, Tatsuki and Nakamura, Kikyo and Sakai, Tadakatsu",
    title = "{Note on higher-group structure in 6d self-dual gauge theory}",
    eprint = "2406.10518",
    archivePrefix = "arXiv",
    primaryClass = "hep-th",
    doi = "10.1007/JHEP10(2024)093",
    journal = "JHEP",
    volume = "10",
    pages = "093",
    year = "2024"
}

@article{Brauner:2020rtz,
    author = "Brauner, Tom{\'a}{\v{s}}",
    title = "{Field theories with higher-group symmetry from composite currents}",
    eprint = "2012.00051",
    archivePrefix = "arXiv",
    primaryClass = "hep-th",
    doi = "10.1007/JHEP04(2021)045",
    journal = "JHEP",
    volume = "04",
    pages = "045",
    year = "2021"
}

@article{Hidaka:2020iaz,
    author = "Hidaka, Yoshimasa and Nitta, Muneto and Yokokura, Ryo",
    title = "{Higher-form symmetries and 3-group in axion electrodynamics}",
    eprint = "2006.12532",
    archivePrefix = "arXiv",
    primaryClass = "hep-th",
    reportNumber = "KEK-TH-2232, J-PARC-TH-0222, RIKEN-iTHEMS-Report-20",
    doi = "10.1016/j.physletb.2020.135672",
    journal = "Phys. Lett. B",
    volume = "808",
    pages = "135672",
    year = "2020"
}

@article{Hidaka:2020izy,
    author = "Hidaka, Yoshimasa and Nitta, Muneto and Yokokura, Ryo",
    title = "{Global 3-group symmetry and 't Hooft anomalies in axion electrodynamics}",
    eprint = "2009.14368",
    archivePrefix = "arXiv",
    primaryClass = "hep-th",
    reportNumber = "KEK-TH-2254, J-PARC-TH-0225, RIKEN-iTHEMS-Report-20",
    doi = "10.1007/JHEP01(2021)173",
    journal = "JHEP",
    volume = "01",
    pages = "173",
    year = "2021"
}

@article{Hidaka:2021mml,
    author = "Hidaka, Yoshimasa and Nitta, Muneto and Yokokura, Ryo",
    title = "{Topological axion electrodynamics and 4-group symmetry}",
    eprint = "2107.08753",
    archivePrefix = "arXiv",
    primaryClass = "hep-th",
    reportNumber = "KEK-TH-2331, J-PARC-TH-0243",
    doi = "10.1016/j.physletb.2021.136762",
    journal = "Phys. Lett. B",
    volume = "823",
    pages = "136762",
    year = "2021"
}

@article{Hidaka:2021kkf,
    author = "Hidaka, Yoshimasa and Nitta, Muneto and Yokokura, Ryo",
    title = "{Global 4-group symmetry and {\textquoteright}t Hooft anomalies in topological axion electrodynamics}",
    eprint = "2108.12564",
    archivePrefix = "arXiv",
    primaryClass = "hep-th",
    reportNumber = "KEK-TH-2346, J-PARC-TH-0249,RIKEN-iTHEMS-Report-21, J-PARC-TH-0249",
    doi = "10.1093/ptep/ptab150",
    journal = "PTEP",
    volume = "2022",
    number = "4",
    pages = "04A109",
    year = "2022"
}

@article{Breen:2001ie,
    author = "Breen, Lawrence and Messing, William",
    title = "{Differential geometry of GERBES}",
    eprint = "math/0106083",
    archivePrefix = "arXiv",
    doi = "10.1016/j.aim.2005.06.014",
    journal = "Adv. Math.",
    volume = "198",
    pages = "732",
    year = "2005"
}

@article{Baez:2004in,
    author = "Baez, John and Schreiber, Urs",
    title = "{Higher gauge theory: 2-connections on 2-bundles}",
    eprint = "hep-th/0412325",
    archivePrefix = "arXiv",
    month = "12",
    year = "2004"
}

@ARTICLE{Schreiber:2007S,
       author = {{Schreiber}, Urs and {Waldorf}, Konrad},
        title = "{Parallel Transport and Functors}",
      journal = {arXiv e-prints},
         year = 2007,
        month = may,
          eid = {arXiv:0705.0452},
        pages = {arXiv:0705.0452},
          doi = {10.48550/arXiv.0705.0452},
archivePrefix = {arXiv},
       eprint = {0705.0452},
 primaryClass = {math.DG},
       adsurl = {https://ui.adsabs.harvard.edu/abs/2007arXiv0705.0452S}
}

@ARTICLE{Schreiber:2008S,
       author = {{Schreiber}, Urs and {Waldorf}, Konrad},
        title = "{Smooth Functors vs. Differential Forms}",
      journal = {arXiv e-prints},
         year = 2008,
        month = feb,
          eid = {arXiv:0802.0663},
        pages = {arXiv:0802.0663},
          doi = {10.48550/arXiv.0802.0663},
archivePrefix = {arXiv},
       eprint = {0802.0663},
 primaryClass = {math.DG},
       adsurl = {https://ui.adsabs.harvard.edu/abs/2008arXiv0802.0663S}
}

@article{Martins:2009evc,
    author = "Martins, Joao Faria and Picken, Roger",
    title = "{The fundamental Gray 3-groupoid of a smooth manifold and local 3-dimensional holonomy based on a 2-crossed module}",
    eprint = "0907.2566",
    archivePrefix = "arXiv",
    primaryClass = "math.CT",
    doi = "10.1016/j.difgeo.2010.10.002",
    month = "7",
    year = "2009"
}

@article{Wang:2013dwa,
    author = "Wang, Wei",
    title = "{On 3-gauge transformations, 3-curvatures, and Gray-categories}",
    eprint = "1311.3796",
    archivePrefix = "arXiv",
    primaryClass = "math-ph",
    doi = "10.1063/1.4870640",
    journal = "J. Math. Phys.",
    volume = "55",
    pages = "043506",
    year = "2014"
}

@article{Saemann:2013pca,
    author = {S{\"a}emann, Christian and Wolf, Martin},
    title = "{Six-Dimensional Superconformal Field Theories from Principal 3-Bundles over Twistor Space}",
    eprint = "1305.4870",
    archivePrefix = "arXiv",
    primaryClass = "hep-th",
    reportNumber = "EMPG-13-07, DMUS-MP-13-12",
    doi = "10.1007/s11005-014-0704-3",
    journal = "Lett. Math. Phys.",
    volume = "104",
    pages = "1147--1188",
    year = "2014"
}

@article{Witten:1998wy,
    author = "Witten, Edward",
    title = "{AdS/CFT correspondence and topological field theory.}",
    eprint = "hep-th/9812012",
    archivePrefix = "arXiv",
    reportNumber = "IASSNS-HEP-98-96",
    doi = "10.1088/1126-6708/1998/12/012",
    journal = "JHEP",
    volume = "12",
    pages = "012",
    year = "1998"
}

@article{Apruzzi:2021nmk,
    author = "Apruzzi, Fabio and Bonetti, Federico and Garc{\'\i}a Etxebarria, I{\~n}aki and Hosseini, Saghar S. and Schafer-Nameki, Sakura",
    title = "{Symmetry TFTs from String Theory}",
    eprint = "2112.02092",
    archivePrefix = "arXiv",
    primaryClass = "hep-th",
    doi = "10.1007/s00220-023-04737-2",
    journal = "Commun. Math. Phys.",
    volume = "402",
    number = "1",
    pages = "895--949",
    year = "2023"
}

@article{Najjar:2024vmm,
    author = "Najjar, Marwan and Santilli, Leonardo and Wang, Yi-Nan",
    title = "{({\ensuremath{-}}1)-form symmetries from M-theory and SymTFTs}",
    eprint = "2411.19683",
    archivePrefix = "arXiv",
    primaryClass = "hep-th",
    reportNumber = "USTC-ICTS/PCFT-24-53",
    doi = "10.1007/JHEP03(2025)134",
    journal = "JHEP",
    volume = "03",
    pages = "134",
    year = "2025"
}

@article{Najjar:2025htp,
    author = "Najjar, Marwan",
    title = "{Modified instanton sum and 4-group structure in 4d $\mathcal{N}=1$$SU(M)$ SYM from holography}",
    eprint = "2503.17108",
    archivePrefix = "arXiv",
    primaryClass = "hep-th",
    month = "3",
    year = "2025"
}

@article{Khlaif:2025jnx,
    author = "Khlaif, Osama and Najjar, Marwan",
    title = "{Aspects of 4d $\mathcal{N}=1$$ADE$ gauge theories from M-theory: decomposition, automorphisms, and generalised symmetries}",
    eprint = "2508.00564",
    archivePrefix = "arXiv",
    primaryClass = "hep-th",
    month = "8",
    year = "2025"
}

@article{Bergman:2025isp,
    author = "Bergman, Oren and Garcia-Valdecasas, Eduardo and Mignosa, Francesco and Rodriguez-Gomez, Diego",
    title = "{The SymTFT of $u(N)$ Yang-Mills Theory and Holography}",
    eprint = "2508.00992",
    archivePrefix = "arXiv",
    primaryClass = "hep-th",
    month = "8",
    year = "2025"
}

@article{Heidenreich:2020pkc,
    author = "Heidenreich, Ben and McNamara, Jacob and Montero, Miguel and Reece, Matthew and Rudelius, Tom and Valenzuela, Irene",
    title = "{Chern-Weil global symmetries and how quantum gravity avoids them}",
    eprint = "2012.00009",
    archivePrefix = "arXiv",
    primaryClass = "hep-th",
    reportNumber = "ACFI-T20-16",
    doi = "10.1007/JHEP11(2021)053",
    journal = "JHEP",
    volume = "11",
    pages = "053",
    year = "2021"
}

@article{Thouless:1982zz,
    author = "Thouless, D. J. and Kohmoto, M. and Nightingale, M. P. and den Nijs, M.",
    title = "{Quantized Hall Conductance in a Two-Dimensional Periodic Potential}",
    doi = "10.1103/PhysRevLett.49.405",
    journal = "Phys. Rev. Lett.",
    volume = "49",
    pages = "405--408",
    year = "1982"
}

@article{Laughlin:1983fy,
    author = "Laughlin, R. B.",
    title = "{Anomalous quantum Hall effect: An Incompressible quantum fluid with fractionallycharged excitations}",
    doi = "10.1103/PhysRevLett.50.1395",
    journal = "Phys. Rev. Lett.",
    volume = "50",
    pages = "1395",
    year = "1983"
}

@article{Wen:1989iv,
    author = "Wen, X. G.",
    title = "{Topological Order in Rigid States}",
    reportNumber = "NSF-ITP-89-107",
    doi = "10.1142/S0217979290000139",
    journal = "Int. J. Mod. Phys. B",
    volume = "4",
    pages = "239",
    year = "1990"
}

@article{Wen:1990zza,
    author = "Wen, X. G. and Niu, Q.",
    title = "{Ground-state degeneracy of the fractional quantum Hall states in the presence of a random potential and on high-genus Riemann surfaces}",
    doi = "10.1103/PhysRevB.41.9377",
    journal = "Phys. Rev. B",
    volume = "41",
    pages = "9377--9396",
    year = "1990"
}

@article{Wen:1991rp,
    author = "Wen, Xiao-Gang",
    title = "{Topological orders and Chern-Simons theory in strongly correlated quantum liquid}",
    reportNumber = "IASSNS-HEP-91-20",
    doi = "10.1142/S0217979291001541",
    journal = "Int. J. Mod. Phys. B",
    volume = "5",
    pages = "1641--1648",
    year = "1991"
}

@article{Nussinov:2006iva,
    author = "Nussinov, Zohar and Ortiz, Gerardo",
    title = "{Sufficient symmetry conditions for Topological Quantum Order}",
    eprint = "cond-mat/0605316",
    archivePrefix = "arXiv",
    doi = "10.1073/pnas.0803726105",
    journal = "Proc. Nat. Acad. Sci.",
    volume = "106",
    pages = "16944--16949",
    year = "2009"
}

@article{Nussinov:2009zz,
    author = "Nussinov, Zohar and Ortiz, Gerardo",
    title = "{A symmetry principle for topological quantum order}",
    eprint = "cond-mat/0702377",
    archivePrefix = "arXiv",
    doi = "10.1016/j.aop.2008.11.002",
    journal = "Annals Phys.",
    volume = "324",
    pages = "977--1057",
    year = "2009"
}

@article{Shao:2023gho,
    author = "Shao, Shu-Heng",
    title = "{What's Done Cannot Be Undone: TASI Lectures on Non-Invertible Symmetries}",
    eprint = "2308.00747",
    archivePrefix = "arXiv",
    primaryClass = "hep-th",
    reportNumber = "YITP-SB-2023-19",
    month = "8",
    year = "2023"
}

@article{Yokokura:2022alv,
    author = "Yokokura, Ryo",
    title = "{Non-invertible symmetries in axion electrodynamics}",
    eprint = "2212.05001",
    archivePrefix = "arXiv",
    primaryClass = "hep-th",
    reportNumber = "KEK-TH-2481",
    month = "12",
    year = "2022"
}

@article{Birmingham:1991ty,
    author = "Birmingham, Danny and Blau, Matthias and Rakowski, Mark and Thompson, George",
    title = "{Topological field theory}",
    reportNumber = "CERN-TH-6045-91",
    doi = "10.1016/0370-1573(91)90117-5",
    journal = "Phys. Rept.",
    volume = "209",
    pages = "129--340",
    year = "1991"
}

@article{Cordova:2018cvg,
    author = "C{\'o}rdova, Clay and Dumitrescu, Thomas T. and Intriligator, Kenneth",
    title = "{Exploring 2-Group Global Symmetries}",
    eprint = "1802.04790",
    archivePrefix = "arXiv",
    primaryClass = "hep-th",
    doi = "10.1007/JHEP02(2019)184",
    journal = "JHEP",
    volume = "02",
    pages = "184",
    year = "2019"
}

@article{Benini:2018reh,
    author = "Benini, Francesco and C{\'o}rdova, Clay and Hsin, Po-Shen",
    title = "{On 2-Group Global Symmetries and their Anomalies}",
    eprint = "1803.09336",
    archivePrefix = "arXiv",
    primaryClass = "hep-th",
    reportNumber = "SISSA 10/2018/FISI, SISSA-10-2018-FISI",
    doi = "10.1007/JHEP03(2019)118",
    journal = "JHEP",
    volume = "03",
    pages = "118",
    year = "2019"
}

@article{Hitchin:1999fh,
    author = "Hitchin, Nigel J.",
    editor = "Vafa, Cumrun and Yau, S. -T.",
    title = "{Lectures on special Lagrangian submanifolds}",
    eprint = "math/9907034",
    archivePrefix = "arXiv",
    journal = "AMS/IP Stud. Adv. Math.",
    volume = "23",
    pages = "151--182",
    year = "2001"
}

@article{Baez:2010ya,
    author = "Baez, John C. and Huerta, John",
    title = "{An Invitation to Higher Gauge Theory}",
    eprint = "1003.4485",
    archivePrefix = "arXiv",
    primaryClass = "hep-th",
    doi = "10.1007/s10714-010-1070-9",
    journal = "Gen. Rel. Grav.",
    volume = "43",
    pages = "2335--2392",
    year = "2011"
}

@article{Schreiber:2008kcv,
    author = "Schreiber, Urs and Waldorf, Konrad",
    title = "{Connections on non-abelian Gerbes and their Holonomy}",
    eprint = "0808.1923",
    archivePrefix = "arXiv",
    primaryClass = "math.DG",
    month = "8",
    year = "2008"
}

@article{Martins:2007uki,
    author = "Martins, Joao Faria and Picken, Roger",
    title = "{On two-Dimensional Holonomy}",
    eprint = "0710.4310",
    archivePrefix = "arXiv",
    primaryClass = "math.DG",
    doi = "10.1090/S0002-9947-2010-04857-3",
    journal = "Trans. Am. Math. Soc.",
    volume = "362",
    pages = "5657--5695",
    year = "2010"
}

@article{MARTINS20113309,
title = {Surface holonomy for non-abelian 2-bundles via double groupoids},
journal = {Advances in Mathematics},
volume = {226},
number = {4},
pages = {3309-3366},
year = {2011},
issn = {0001-8708},
doi = {https://doi.org/10.1016/j.aim.2010.10.017},
url = {https://www.sciencedirect.com/science/article/pii/S0001870810003804},
author = {João Faria Martins and Roger Picken}
}

@article{Horowitz:1989ng,
    author = "Horowitz, Gary T.",
    title = "{Exactly Soluble Diffeomorphism Invariant Theories}",
    reportNumber = "NSF-ITP-88-178",
    doi = "10.1007/BF01218410",
    journal = "Commun. Math. Phys.",
    volume = "125",
    pages = "417",
    year = "1989"
}

@article{Blau:1989bq,
    author = "Blau, Matthias and Thompson, George",
    title = "{Topological Gauge Theories of Antisymmetric Tensor Fields}",
    reportNumber = "SISSA-39/89/FM, PAR-LPTHE-89-17",
    doi = "10.1016/0003-4916(91)90240-9",
    journal = "Annals Phys.",
    volume = "205",
    pages = "130--172",
    year = "1991"
}

@article{Blau:1989dh,
    author = "Blau, Matthias and Thompson, George",
    title = "{A New Class of Topological Field Theories and the Ray-singer Torsion}",
    reportNumber = "PAR-LPTHE-89-18",
    doi = "10.1016/0370-2693(89)90526-1",
    journal = "Phys. Lett. B",
    volume = "228",
    pages = "64--68",
    year = "1989"
}

@article{Kapustin:2013uxa,
    author = "Kapustin, Anton and Thorngren, Ryan",
    title = "{Higher Symmetry and Gapped Phases of Gauge Theories}",
    eprint = "1309.4721",
    archivePrefix = "arXiv",
    primaryClass = "hep-th",
    doi = "10.1007/978-3-319-59939-7_5",
    journal = "Prog. Math.",
    volume = "324",
    pages = "177--202",
    year = "2017"
}

@article{Sharpe:2015mja,
    author = "Sharpe, Eric",
    title = "{Notes on generalized global symmetries in QFT}",
    eprint = "1508.04770",
    archivePrefix = "arXiv",
    primaryClass = "hep-th",
    doi = "10.1002/prop.201500048",
    journal = "Fortsch. Phys.",
    volume = "63",
    pages = "659--682",
    year = "2015"
}

@article{Girelli:2007tt,
    author = "Girelli, F. and Pfeiffer, H. and Popescu, E. M.",
    title = "{Topological Higher Gauge Theory - from BF to BFCG theory}",
    eprint = "0708.3051",
    archivePrefix = "arXiv",
    primaryClass = "hep-th",
    doi = "10.1063/1.2888764",
    journal = "J. Math. Phys.",
    volume = "49",
    pages = "032503",
    year = "2008"
}

@article{Radenkovic:2019qme,
    author = "Radenkovic, Tijana and Vojinovic, Marko",
    title = "{Higher Gauge Theories Based on 3-groups}",
    eprint = "1904.07566",
    archivePrefix = "arXiv",
    primaryClass = "hep-th",
    doi = "10.1007/JHEP10(2019)222",
    journal = "JHEP",
    volume = "10",
    pages = "222",
    year = "2019"
}

@article{Stueckelberg:1957zz,
	author = {Stueckelberg, E. C. G.},
	doi = {10.5169/seals-112814},
	journal = {Helv. Phys. Acta},
	pages = {209-215},
	slaccitation = {%%CITATION = HPACA,30,209;%%},
	title = {{Theory of the radiation of photons of small arbitrary mass}},
	volume = {30},
	year = {1957}
}

@article{McGreevy:2022oyu,
    author = "McGreevy, John",
    title = "{Generalized Symmetries in Condensed Matter}",
    eprint = "2204.03045",
    archivePrefix = "arXiv",
    primaryClass = "cond-mat.str-el",
    doi = "10.1146/annurev-conmatphys-040721-021029",
    journal = "Ann. Rev. Condensed Matter Phys.",
    volume = "14",
    pages = "57--82",
    year = "2023"
}

@inproceedings{Cordova:2022ruw,
	archiveprefix = {arXiv},
	author = {Cordova, Clay and Dumitrescu, Thomas T. and Intriligator, Kenneth and Shao, Shu-Heng},
	booktitle = {{Snowmass 2021}},
	eprint = {2205.09545},
	month = {5},
	primaryclass = {hep-th},
	title = {{Snowmass White Paper: Generalized Symmetries in Quantum Field Theory and Beyond}},
	year = {2022}
}

@article{Brennan:2023mmt,
	archiveprefix = {arXiv},
	author = {Brennan, T. Daniel and Hong, Sungwoo},
	eprint = {2306.00912},
	month = {6},
	primaryclass = {hep-ph},
	title = {{Introduction to Generalized Global Symmetries in QFT and Particle Physics}},
	year = {2023}
}

@article{Luo:2023ive,
    author = "Luo, Ran and Wang, Qing-Rui and Wang, Yi-Nan",
    title = "{Lecture notes on generalized symmetries and applications}",
    eprint = "2307.09215",
    archivePrefix = "arXiv",
    primaryClass = "hep-th",
    doi = "10.1016/j.physrep.2024.02.002",
    journal = "Phys. Rept.",
    volume = "1065",
    pages = "1--43",
    year = "2024"
}

@article{Gomes:2023ahz,
	archiveprefix = {arXiv},
	author = {Gomes, Pedro R. S.},
	doi = {10.21468/SciPostPhysLectNotes.74},
	eprint = {2303.01817},
	journal = {SciPost Phys. Lect. Notes},
	pages = {1},
	primaryclass = {hep-th},
	title = {{An introduction to higher-form symmetries}},
	volume = {74},
	year = {2023}
}

@article{DelZotto:2024ngj,
    author = "Del Zotto, Michele and Dell'Acqua, Matteo and Riedel G{\r{a}}rding, Elias",
    title = "{The Higher Structure of Symmetries of Axion-Maxwell Theory}",
    eprint = "2411.09685",
    archivePrefix = "arXiv",
    primaryClass = "hep-th",
    month = "11",
    year = "2024"
}

@article{Copetti:2023mcq,
    author = "Copetti, Christian and Del Zotto, Michele and Ohmori, Kantaro and Wang, Yifan",
    title = "{Higher Structure of Chiral Symmetry}",
    eprint = "2305.18282",
    archivePrefix = "arXiv",
    primaryClass = "hep-th",
    doi = "10.1007/s00220-024-05227-9",
    journal = "Commun. Math. Phys.",
    volume = "406",
    number = "4",
    pages = "73",
    year = "2025"
}

@article{Choi:2022fgx,
    author = "Choi, Yichul and Lam, Ho Tat and Shao, Shu-Heng",
    title = "{Non-invertible Gauss law and axions}",
    eprint = "2212.04499",
    archivePrefix = "arXiv",
    primaryClass = "hep-th",
    reportNumber = "MIT-CTP/5504, YITP-SB-2022-39",
    doi = "10.1007/JHEP09(2023)067",
    journal = "JHEP",
    volume = "09",
    pages = "067",
    year = "2023"
}

@article{zbMATH07940412,
    author = {Sarikaya, Murat and Ulualan, Erdal},
    title = {Comparing 2-crossed modules with {Gray} 3-groups},
    fjournal = {Theory and Applications of Categories},
    journal = {Theory Appl. Categ.},
    issn = {1201-561X},
    volume = {41},
    pages = {1557--1595},
    year = {2024},
    language = {English},
    url = {www.tac.mta.ca/tac/volumes/41/45/41-45abs.html#vol41},
    zbMATH = {7940412},
    Zbl = {1550.18003}
}

@article{zbMATH05664819,
    author = {Arvasi, Z. and Kuzpinari, T. S. and Uslu, E. {\"O}.},
    title = {Three-crossed modules},
    fjournal = {Homology, Homotopy and Applications},
    journal = {Homology Homotopy Appl.},
    issn = {1532-0073},
    volume = {11},
    number = {2},
    pages = {161--187},
    year = {2009},
    language = {English},
    doi = {10.4310/HHA.2009.v11.n2.a8},
    eprint={0812.4685},
    archivePrefix={arXiv},
    primaryClass={math.CT},
    url = {intlpress.com/hha/v11/n2/a8/pdf}, 
}

@article{Fukuda2025,
    title="{3-Crossed modules, Quasi-categories, and the Moore complex}", 
    author={Masaki Fukuda and Tommy Shu},
    year={2025},
    eprint={2512.22797},
    archivePrefix={arXiv},
    primaryClass={math.CT},
    url={https://arxiv.org/abs/2512.22797}, 
}

@unpublished{Fukuda:2026,
    author  = {Masaki Fukuda and Tommy Shu},
    title   = "{3-crossed modules and generalizations of Gray-categories}",
    note = {in preparation},
    year    = {2026}
}

\end{document}